\documentclass[preprint,onecolumn,nobibnotes,showpacs,preprintnumbers,superscriptaddress,amsmath,amssymb]{revtex4}
\pdfoutput=1
\usepackage{booktabs,graphicx,pstricks,verbatim,url}
\newcommand{\kev}{\ensuremath {\rm keV}}

\newcommand{\gev}{\ensuremath {\rm GeV}}

\newcommand{\be}{\begin{equation}}
\newcommand{\ee}{\end{equation}}
\newcommand{\leff}{\ensuremath{{\cal L}_{eff}}}
\newcommand{\ely}{\ensuremath{{\epsilon_{LY}}}}

\usepackage{accents}
\newlength{\dhatheight}

\def\bea{\begin{eqnarray}}
\def\eea{\end{eqnarray}}
\def\ltap{\ \raise.3ex\hbox{$<$\kern-.75em\lower1ex\hbox{$\sim$}}\ }
\def\gtap{\ \raise.3ex\hbox{$>$\kern-.75em\lower1ex\hbox{$\sim$}}\ }
\def\lsim{\ \raise.3ex\hbox{$<$\kern-.75em\lower1ex\hbox{$\sim$}}\ }
\def\gsim{\ \raise.3ex\hbox{$>$\kern-.75em\lower1ex\hbox{$\sim$}}\ }

\newcommand{\vmin}{v_{\mathrm{min}}}
\newcommand{\vesc}{v_{\mathrm{esc}}}
\usepackage{color}
\begin{document}
\begin{flushright}
FERMILAB-PUB-13-583-T
\end{flushright}

\title{Dark Matter in Light of LUX}
\author{Patrick J. Fox}
\affiliation{Theoretical Physics Department, Fermilab, Batavia, Illinois 60510, USA}
\author{Gabriel Jung}
\affiliation{Theoretical Physics Department, Fermilab, Batavia, Illinois 60510, USA}
\author{Peter Sorensen}
\affiliation{Lawrence Livermore National Laboratory, 7000 East Ave., Livermore, CA 94550, USA}
\author{Neal Weiner}
\affiliation{Center for Cosmology and Particle Physics,  Department of Physics, New York University, New York, NY 10003, USA}
\begin{abstract}
The landscape of dark matter direct detection has been profoundly altered by the slew of recent experiments. While some have claimed signals consistent with dark matter, others have seen few, if any, events consistent with dark matter. The results of the putative detections are often incompatible with each other in the context of naive spin-independent scattering, as well as with the null results. In particular, in light of the conflicts between the DM interpretation of the three events recently reported by the CDMS-Si experiment and the first results of the LUX experiment, there is a strong need to revisit the assumptions that go into the DM interpretations of both signals and limits. We attempt to reexamine a number of particle physics, astrophysics and experimental uncertainties. Specifically, we examine exothermic scattering, isospin-dependent couplings, modified halo models through astrophysics independent techniques, and variations in the assumptions about the scintillation light in liquid Xenon.  We find that only a highly tuned isospin-dependent scenario remains as a viable explanation of the claimed detections, unless the scintillation properties of LXe are dramatically different from the assumptions used by the LUX experiment.
\end{abstract}
\maketitle
\section{Introduction}\label{sec:intro}

Since first proposed several decades ago \cite{Goodman:1984dc}, the quest to directly observe dark matter (DM) interacting with the standard model (SM) has grown to include a multitude of experiments, employing varying techniques and technologies, to discern a tiny DM signal from amongst potentially large backgrounds.  The bounds on the interaction strength of DM have improved by many orders of magnitude, both for DM coupling in a spin-dependent or spin-independent fashion to the nucleus.  For spin-dependent the leading constraints come from SIMPLE~\cite{Felizardo:2011uw}, PICASSO~\cite{Archambault:2012pm}, and COUPP~\cite{Behnke:2012ys} and for spin-independent from CDMS-Ge~\cite{Ahmed:2009zw,Ahmed:2010wy}, XENON100~\cite{Aprile:2012nq}, and most recently LUX~\cite{Akerib:2013tjd}.  At the same time, some experiments have found anomalies that may be due to a DM signal, particularly in regions consistent with a low mass ($m_\chi\ltap 10$ GeV).  Most recently, CDMS-Si~\cite{Agnese:2013rvf} has seen 3 recoil events, a rate that would be expected from the known backgrounds only $\sim 0.2\%$ of the time.  This joins the longer standing discrepancies of DAMA~\cite{Bernabei:2010mq}, CoGeNT~\cite{Aalseth:2010vx}, and CRESST~\cite{Angloher:2011uu}.  Interpreted as elastically scattering DM coupled in a spin-independent fashion to nuclei, and assuming the standard halo model (SHM) for the distribution of DM speeds, these two most recent results, CDMS-Si and LUX, appear to be in considerable tension with one another~\cite{Gresham:2013mua,DelNobile:2013gba,
Cirigliano:2013zta}.  In this paper, we wish investigate the robustness of this statement.  

In particular, there are several aspects of liquid noble detectors, such as LUX, that have recently received renewed scrutiny.  We consider a broad range for the parameters characterizing important detector responses and investigate the impact they have on the tensions.
\begin{itemize}
\item \textbf{Light yield of xenon in an electric field}  The LUX experiment operates in a background electric field of 181 V/cm. The scintillation response of xenon to nuclear recoils has not been conclusively measured, in particular at low energies, in such a high electric field. It it is known for argon \cite{Alexander:2013aaq} that the electric field can alter the S1 light yield by at least 10\%-20\%, leading to a suppression factor $\epsilon_{\mathrm{LY}}\sim 0.9-0.8$.  Simulations indicate the reduction are around 20\% for the LUX running conditions~\cite{Szydagis:2013sih}.  However, this has not been measured, and the sensitivity of these conclusions to this suppression warrants further examination. In this work, we consider a broad range, $\epsilon_{\mathrm{LY}}\in [0.2,0.8]$.
\item \textbf{Scintillation efficiency ($\mathcal{L}_{eff}$)} Similarly the scintillation efficiency ($\mathcal{L}_{eff}$) is a subject of much debate.  This has not been measured below nuclear recoil energies of 3 keV, although it is believed to be non-zero, but decreasing, as one goes to lower recoil energies\cite{Sorensen:2011bd}.  Again, we will remain agnostic and instead consider the effects of cutting off the extrapolation below 3 keV.
\item \textbf{Poisson fluctuations} The dominant sensitivity of liquid xenon experiments to light WIMPs arises from upward fluctuations in the $S_1$ signal, which generally has an expectation value below the $S_1 = 2 {\rm P.E.}$ threshold.  Thus it is not sufficient to determine the average amount of scintillation light for a given recoil, but instead to take into account the fluctuations.  We take into account these, and other, fluctuations in how the detector responds to a nuclear recoil when modeling the LUX response, see below for details.
\end{itemize}

Just as the experimental DM program has grown over the years so has the theoretical program.  Our understanding of the possibilities for DM phenomenology and the character of its interactions is no longer limited to those of a neutralino.  We consider the leading three ways to try to remove the tension between the excess at CDMS-Si and the null results of LUX.
\begin{itemize}
\item \textbf{Deviations from the SHM}  Although N-body simulations provide considerable evidence that on large scales the DM halo has an approximately Maxwellian velocity distribution it is unknown what its form is on solar system scales.  Thus, it may be that the tension is due to the assumption that the distribution probed by both experiments is Maxwellian in nature. In particular, for very light WIMPs ($m_\chi \lsim 6\ \gev$) the events at CDMS are probing the very tail of the velocity distribution where deviations from a smooth halo are most likely to appear. We will compare the results of the two experiments in an astrophysics independent fashion\cite{Fox:2010bz,Frandsen:2011gi,Gondolo:2012rs}.
\item \textbf{Isospin dependent couplings} The two experiments are built from different materials with different ratios of protons and neutrons in their nuclei.  If the DM couples with the opposite sign to protons and neutrons, and in the appropriate ratio~\cite{Kurylov:2003ra,Feng:2011vu}, it is possible that the couplings to xenon are considerably suppressed, thus relieving some of the tension.  We will investigate what level on cancellation is necessary to achieve accord.
\item \textbf{Inelastical scattering} If the DM coupling to nuclei occurs through a transition to a nearly degenerate dark state, the kinematics are drastically altered.  Since the excess is seen in an experiment with a light target the preferred kinematics are for the second state to be lighter than the DM, exothermic DM~\cite{Graham:2010ca}. We study the degree to which this can alleviate tensions in the data.
\end{itemize}

This paper is laid out as follows: in section \ref{sec:models} we review the standard assumptions and review the experimental results from LUX as well as the CDMS Ge and Si results. In section \ref{sec:astroindependent} we review the approach of \cite{Fox:2010bz} in approaching dark matter results in a conservative fashion that removes the astrophysical uncertainties. In section \ref{sec:results} we study the existing data with these variations in mind to see what, if any, parameter space is allowed.

\section{DM and Astrophysics models}\label{sec:models}

For DM scattering off nuclei the differential rate is given by,
\be\label{eq:mastereq}
\frac{dR}{dE_R} = \frac{N_T \rho_\chi}{m_\chi}\int_{\vmin}^{\vesc}d^3\vec{v} f(\vec{v}(t))\frac{d\sigma|\vec{v}|}{dE_R},
\ee 
where $N_T$ denotes the number of scattering targets and $\rho_\chi$ is the local DM density (typically taken to be 0.3 GeV/cm$^3$).  We will focus our attentions on spin-independent couplings\footnote{For a possible spin-dependent explanation of the results, see \cite{Buckley:2013gjo}.}, the most abundant isotope of silicon carries no spin, for which the nuclear differential scattering cross section is related to the neutron cross section by,
\be\label{eq:diffrate}
\frac{d\sigma}{dE_R} = F_N^2(E_R) \frac{m_N }{2\mu_{n\chi}^2 v^2} \frac{(Z f_p + (A-Z) f_n)^2}{f_n^2}\sigma_n~.
\ee
The nuclear form factor, $F_N$, takes into account the fact that at non-zero momentum exchange the interaction can resolve the nuclear structure.

In the general case where the scattering of the DM involves a transition of the DM to another dark-sector particle, DM$^\prime$, whose mass differs by $\delta$ the relationship between the minimum incoming speed necessary and recoil energy, $E_R$, is given by, 
\be\label{eq:vmin}
\vmin=\frac{1}{\sqrt{2m_N E_R}}\left|\frac{m_N E_R}{\mu_{N\chi}}+\delta\right|~.
\ee
We now briefly describe the experiments we are interested in.
\subsection*{CDMS-Si}

Using 8 silicon detectors, CDMS collected data between July 2007 and September 2008 corresponding to a total of 140.2 kg-days of exposure, observing 3 nuclear recoil events (with energies of $(8.2,\, 9.5,\, 12.3)$ keV) with a total expected background of 0.7 events~\cite{Agnese:2013rvf}.  Assuming that all backgrounds are known the probability to 
see 3 or more events is $\sim 5\%$, and less if the energy distribution of the events and the backgrounds are taken in to account.  Some alternatives to this background fluctuation include
: none are background, or 2 are signal and 1 is background.  Motivated by the canonical DM spectrum of an exponential, and also by the need to remove tensions with other experiments, we will consider that, in the later case, it is the highest energy event which is a background event.  Furthermore, it is possible that the CMDS experiment was ``lucky" and the 2 or 3 signal events are in fact an upward fluctuation on a smaller expected rate.  At 90\% C.L. these expected rates would be 0.5 and 1.1, respectively.  Although each detector at CDMS has a different efficiency we take the efficiency as the cumulative efficiency curve, presented in \cite{Agnese:2013rvf}.

\subsection*{LUX}

The LUX collaboration \cite{Akerib:2013tjd} has collected $118\times 85.3$ kg-days of data, observing 1 candidate event in the signal region.  In a similar fashion to the CDMS-Si result being an upward fluctuation we will often take the conservative assumption that this event was a downward fluctuation from a higher expected rate, which at 90\% C.L. would be 3.9.  In comparing the two experiments we will often compare these two 90\% C.L.'s to be very conservative.

There has been considerable debate about the energy reconstruction in liquid xenon detectors, and the related efficiency of the detector to a DM recoil event.  Furthermore, many of the measurements of the response of liquid xenon to nuclear recoils has taken place without the, or with smaller, electric field than is present in LUX.  The presence of an electric field reduces the light yield by an energy dependent factor.  This energy dependence is thought to be weak and we model the reduction in light yield from the case of zero electric field as a simple rescaling by an amount $\epsilon_{\mathrm{LY}}<1$.  For LUX we take this to be $\epsilon_{\mathrm{LY}}=0.8$, and we investigate the effect of taking $\epsilon_{\mathrm{LY}}=0.4,\,0.2$.

\begin{figure}[t] 
   \centering
   \includegraphics[width=0.45\columnwidth]{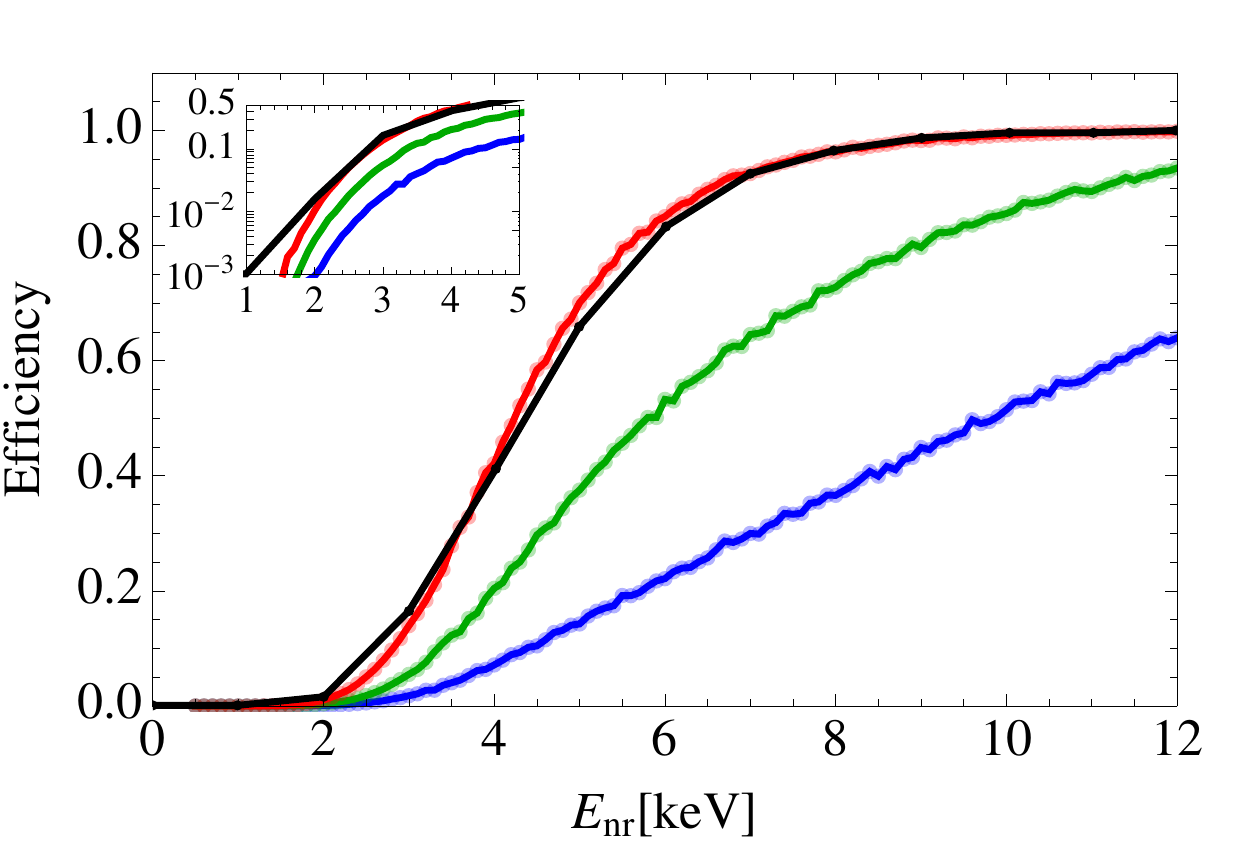} 
    \hspace{0.05\columnwidth}
    \includegraphics[width=0.45\columnwidth]{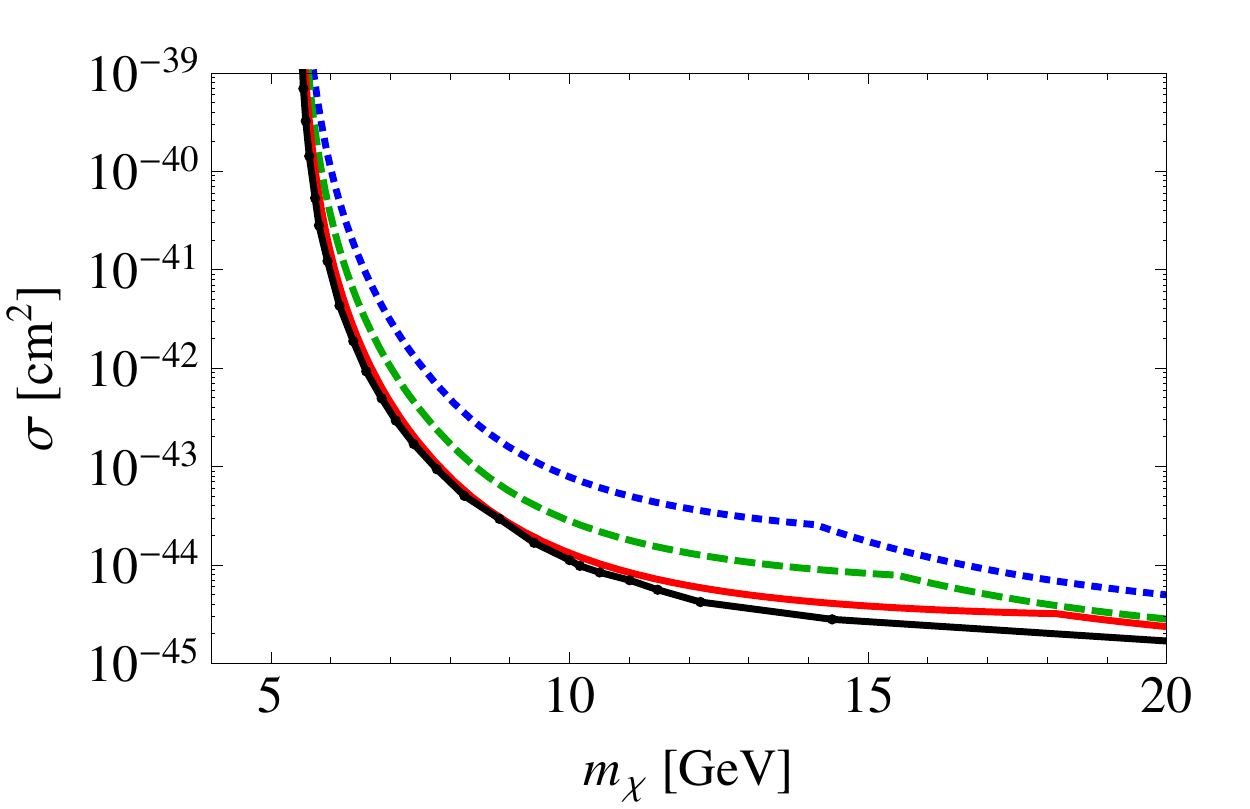} 
   \caption{LH plot: efficiencies at a LUX-like experiment, along with the efficiency presented by LUX (black points) with a linear interpolation.  The red (upper) curve uses an energy-independent reduced light yield, due to the presence of an external electric field, of $\epsilon_{\mathrm{LY}}=0.8$ while the green (middle) has $\epsilon_{\mathrm{LY}}=0.4$ and the blue (lower) curve has $\epsilon_{\mathrm{LY}}=0.2$.  RH plot: the bound on the DM-nucleon cross section for each efficiency, the black curve is the bound presented by LUX.}
   \label{fig:LUXeffs}
\end{figure}

To do so we model the behavior of the LUX detector using the technique described in \cite{Sorensen:2011bd}, which is based on Lindhard theory with $k=0.11$, to determine the S1 and S2 signals.  With $\mathcal{L}_{eff}$ and $\mathcal{Q}_y$ as determined in \cite{Sorensen:2011bd} we can convert a nuclear recoil energy into a mean expected S1 and S2 signal.  Taking into account Poisson fluctuations around these means, the response of the PMTs and analysis cuts of $S1>2$ P.E. (with the requirement that at least 2 PMTs detect more than 0.25 P.E. within 100ns of each other) and $S2>200$ P.E.\footnote{We ignore the requirement that $\log_{10} S2/S1$ lies in the bottom half of the nuclear recoil band.  For the mass range we are interested in this is a small correction.} we can determine the efficiency to see nuclear recoils.  The results of this simulation for $\epsilon_{\mathrm{LY}}=0.8,\,0.4,\,0.2$, along with the efficiency presented in \cite{Akerib:2013tjd} are shown Figure~\ref{fig:LUXeffs}.  Our simulation shows excellent agreement with the results presented by LUX, Figure~\ref{fig:LUXeffs}.  The results presented in \cite{Akerib:2013tjd} make the conservative assumption that the efficiency goes to zero below recoil energy of 3 keV.  We investigate how the bounds are strengthened if the efficiency has the more physically reasonable behaviour shown in Figure~\ref{fig:LUXeffs}, although we will not consider recoil energies below 2 keV.

Using our simulation of the detector it is also possible to determine the most likely energy of the event that passes the cuts.  We simulate the distribution of $(S1,S2)$ for a given nuclear recoil energy, fit this distribution to a multivariate gaussian, and then find the distribution for which the observed event is closest to the mean.   Since LUX only presents $S1$ and $S2_b$ for the observed event we assume the full $S2$ signal is twice as large as $S2_b$, thus the observed event has $(S1,S2)\approx(3.2,360)$.  We estimate the most likely recoil energy for the observed event is $\sim 6$ keV, for all three versions of the efficiency.  

\section{Astroindependent techniques}\label{sec:astroindependent}

The differential scattering rate (\ref{eq:mastereq}) can be written as the product of a detector independent function and a detector dependent term.  Thus, any direct detection result can be reinterpreted as an observation of, or constraint on, the detector independent quantity,
\be
\tilde{g}(\vmin)=\frac{\rho \sigma_p}{m_\chi} g(\vmin)\,
\ee
where $g(\vmin)=\int_{\vmin}^{\vesc} d^3v f(v)/v$ is the dependence of the rate on the integral of the DM velocity distribution.  To compare these detector independent quantities between experiments an assumption has to be made about the DM mass.  Equivalently, by going through this $\vmin$-space \cite{Fox:2010bz} and comparing $\tilde{g}(\vmin)$ it is possible to compare the results of two experiments without making assumptions about the local DM velocity distribution, a poorly determined quantity \cite{Green:2011bv}.  When dealing with actual experimental data there are additional issues one must contend with, such as binning of data, experimental resolution, and efficiencies \cite{Gondolo:2012rs}.  However, due to the low statistics of present searches these are often subleading effects.

For elastically scattering DM there is a one-to-one relationship between the observed energy of an event and the minimum allowed speed the incoming DM could posses.  For inelastic or exothermic DM this relationship is no longer so straightforward.  For any given energy there is still only one $\vmin$, but the inverse is no longer true.  There are two branches, and for a particular $\vmin$ there are two recoil energies this could correspond to.  For energies less than $E_{\mathrm{tp}}=|\mu\delta/m_N|$ the minimum velocity, $\vmin$, is a decreasing function of $E_R$ and for $E_R>E_{\mathrm{tp}}$ it is an increasing function.  Note that at this turning point, $E_{\mathrm{tp}}$, the minimum dark matter speed probed is non-zero for inelastic DM and is 0 km/s for exothermic.  This is as expected since for inelastic the DM must carry sufficient energy to up scattering, while a down scatter can occur for any incoming speed, even zero.

This surjective map from $\vmin$-space to energy space still allows for a simple comparison of multiple experiments by mapping the data from energy to $\vmin$-space, as before.  It also has the interesting feature that a \emph{single} experiment, should it be sensitive to recoil energy both above and below $E_0$, has the capability to probe the same part of velocity space twice.  This double determination of $g(\vmin)$ by a single experiment allows, if there is sufficient data and backgrounds are under control, one to determine whether a putative signal is consistent with a single DM species or not.

\subsection*{The velocity distribution}

Although we will focus on analyzing the data without making assumptions about the form of the DM velocity distribution, $f(\vec{v}(t))$, it is instructive to compare the results against those derived assuming the standard halo model (SHM).  There is considerable uncertainty as to the exact form of the DM phase space distribution in our vicinity of the halo \cite{Green:2011bv}.  We take a simple Maxwell-Boltzmann form for the velocity distribution,
\be\label{eq:MB}
f(v)\propto e^{-v^2/v_0^2}-e^{-\vesc^2/v_0^2}~,
\ee
with $v_0=220$ km/s, $\vesc=550$ km/s, and we take the earth's speed to be $230$ km/s.

Simulations of the evolution of DM halos point towards departures from a pure MB velocity distribution.  We have checked that using the empirical model of \cite{Mao:2012hf}, determined by fitting to the simulations, does not significantly alter our results.

\begin{figure}[t] 
   \centering
   \includegraphics[width=0.45\columnwidth]{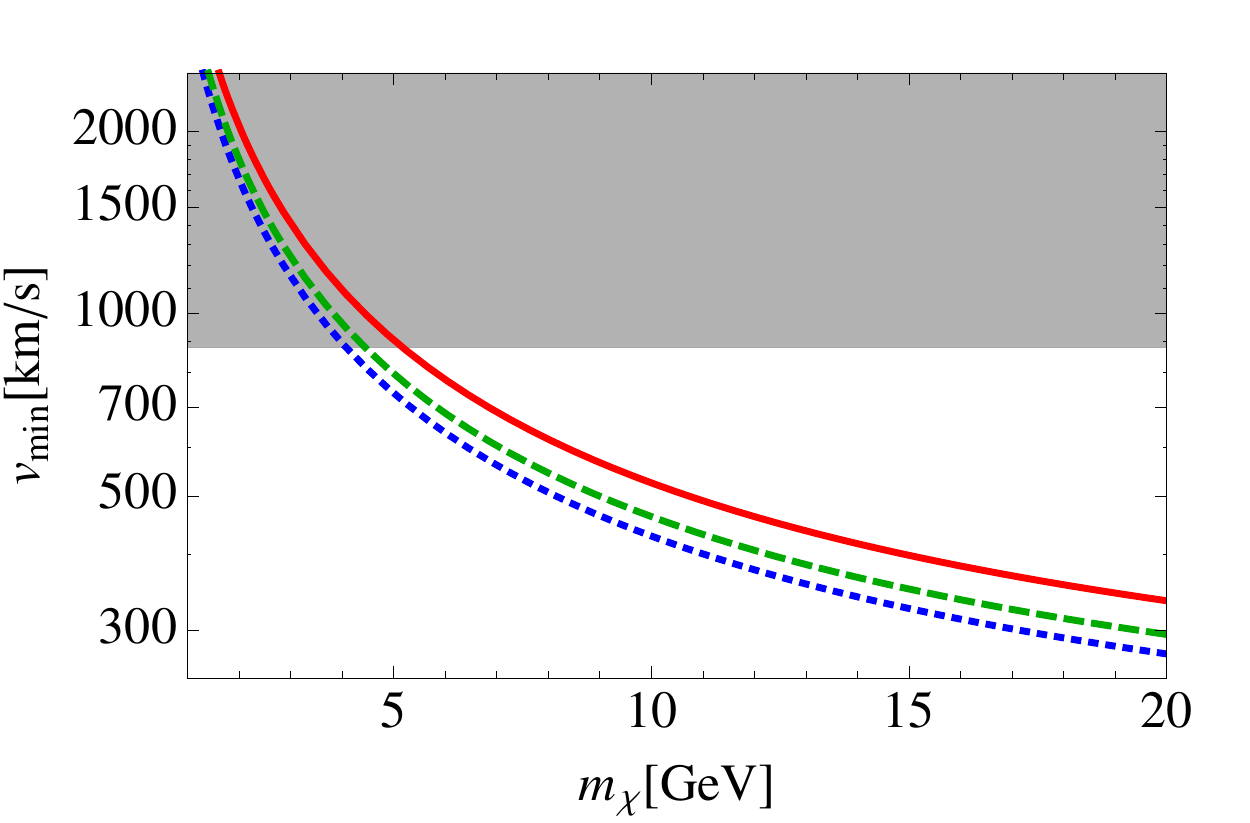} 
    \hspace{0.05\columnwidth}
   \includegraphics[width=0.45\columnwidth]{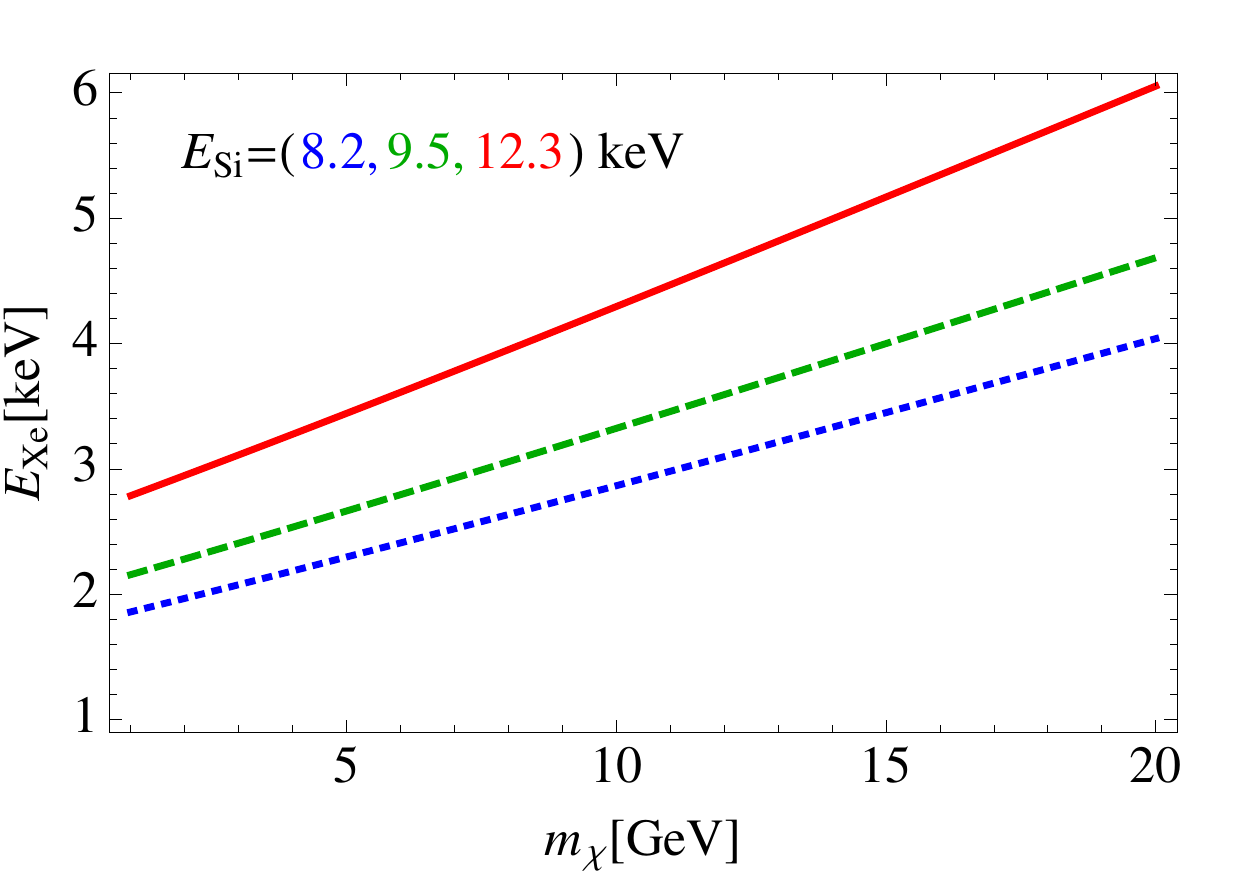}
   \caption{LH plot: the minimum speed the DM must be moving, in the lab frame, to produce a recoil event at each of the energies observed at CDMS-Si.  The upper (red,solid) curve corresponds to 12.29 keV, the middle (green, dashed) to 9.51 keV and the lower (blue, dotted) to 8.2 keV.  RH plot: the recoil energy in xenon for each of the three CDMS-Si events.}
   \label{fig:vminplot}
\end{figure}

\begin{figure}[t] 
   \centering
   \includegraphics[width=0.45\columnwidth]{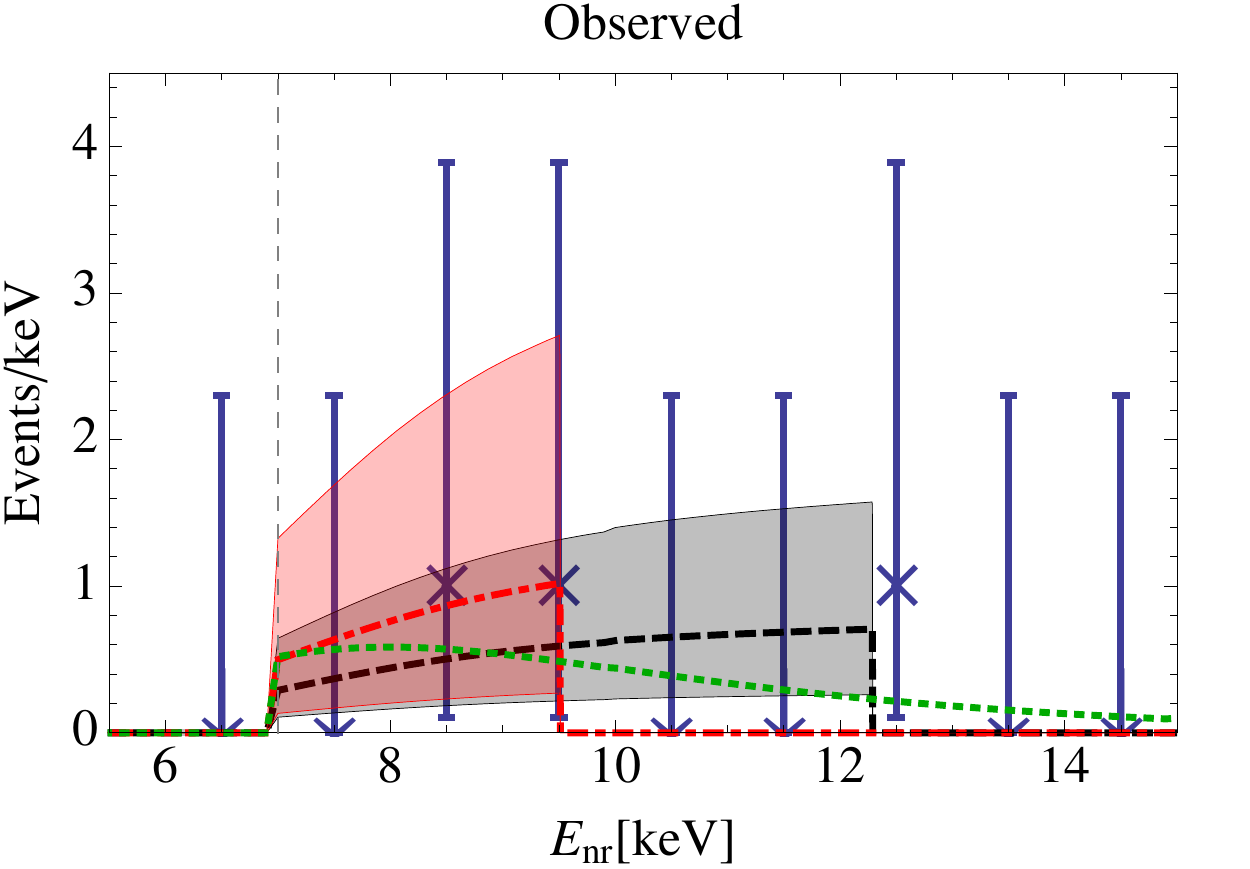}
   \hspace{0.05\columnwidth}
   \includegraphics[width=0.45\columnwidth]{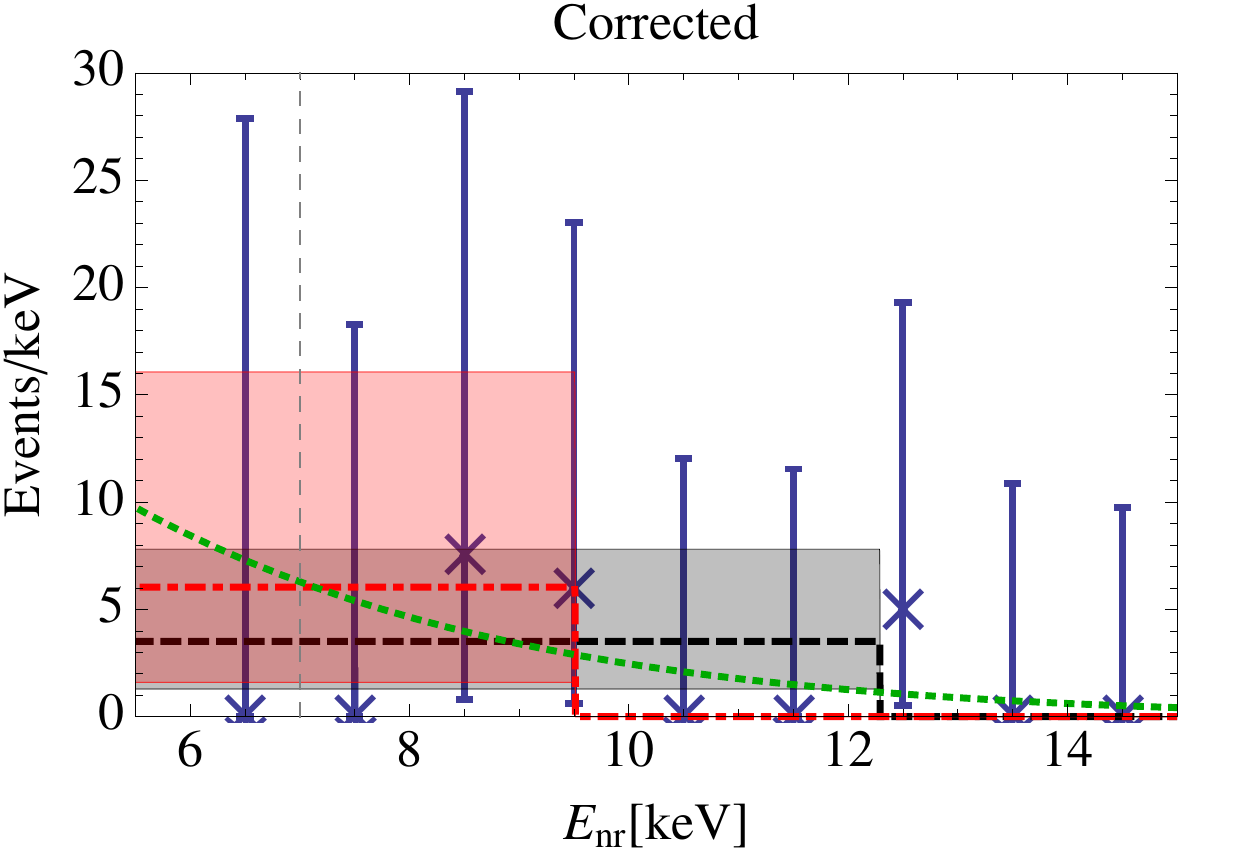} 
   \caption{The event rate at CDMS, in bins of 1 keV.  The error bars denote the range, at 90\% C.L., that is consistent with the number of observed events in each bin.  The RH plot has corrected for the efficiency and form factor, and shows the underlying scattering rate.  The results are consistent with the most conservative assumption of an underlying flat (in energy) event rate.  The black (dashed) curve shows the normalization of the rate, if all three events are taken as signal, such that 3 events would be expected at CDMS-Si.  The red (dot-dashed) curve shows the same if only the two lowest energy events are due to DM.  The surrounding bands in each case are the 90\% C.L. upper and lower limits on the rate.  The green (dotted) curve shows the rate expected for DM of mass 10 GeV if the velocity distribution is MB, as defined in (\ref{eq:MB}).}
   \label{fig:CDMSevents}
\end{figure}
\section{Results}
\label{sec:results}
Using the techniques outlined above we now proceed to compare the constraints from LUX to the results of CDMS-Si, interpreted as WIMP signal.  We investigate the affect upon these constraints from varying the analysis threshold at LUX and the extrapolation of the detector behaviour from it measured response at $E_{nr}\ge 3$ keV with zero background electric field to lower recoil energies and to non-zero background electric fields.  We also show that other CDMS runs, CDMS-Ge with a low threshold and existing data below analysis thresholds can have an impact on this interpretation.

\subsection{eDM}

\begin{figure}[h] 
   \centering
   \includegraphics[width=0.33\columnwidth]{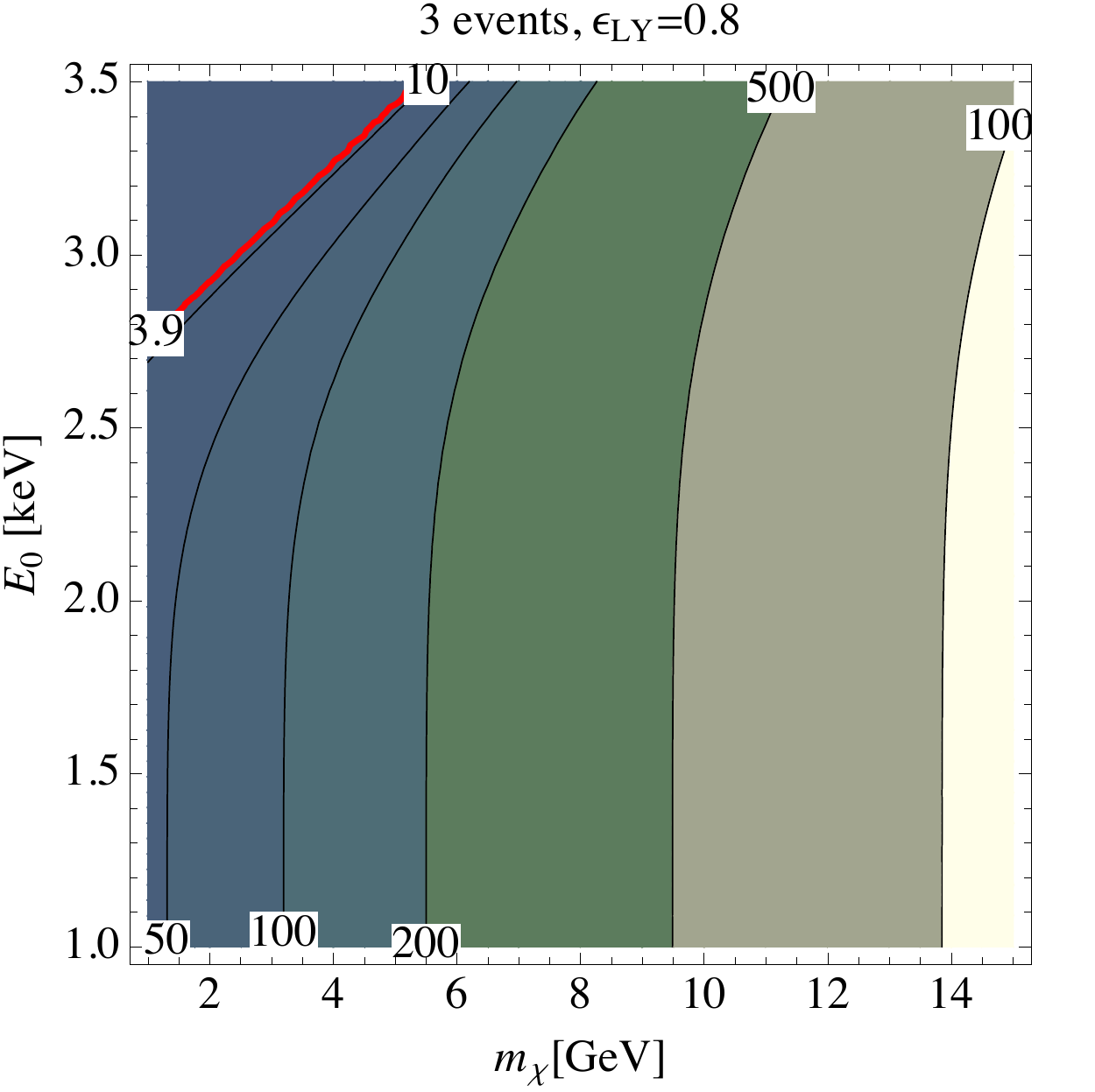}
   \hspace{0.05\columnwidth}
   \includegraphics[width=0.33\columnwidth]{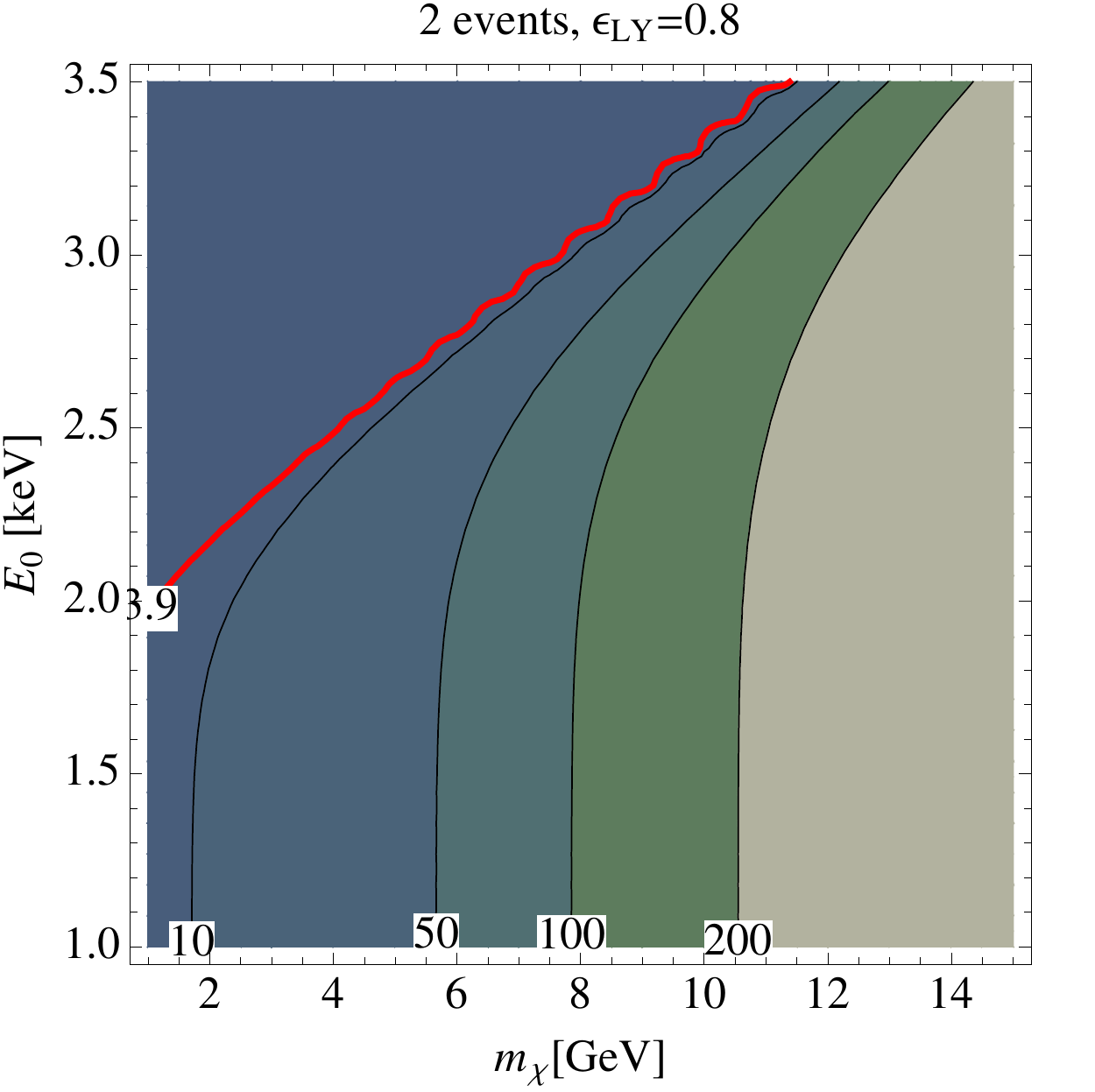} 
   \\
   \includegraphics[width=0.33\columnwidth]{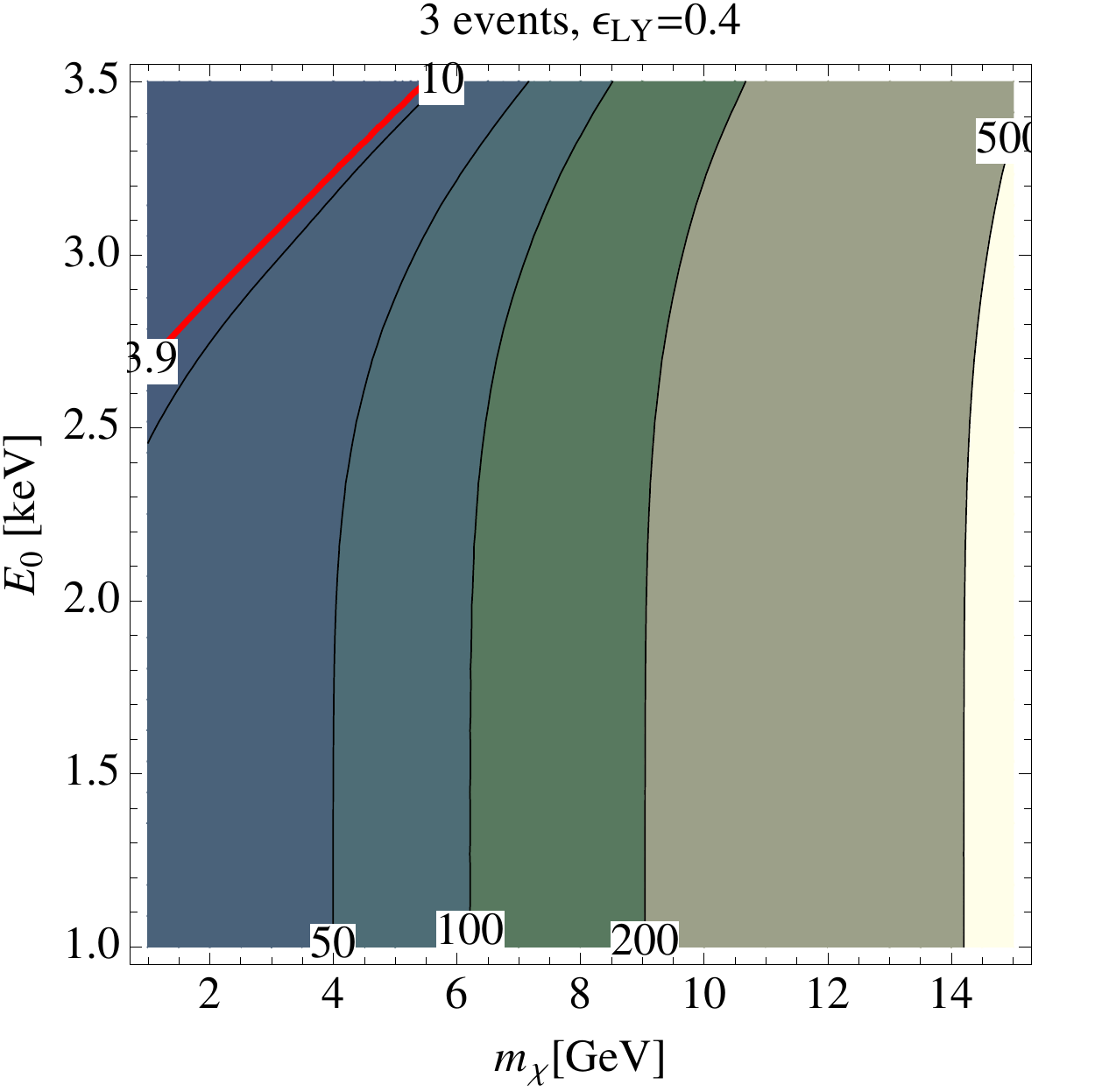}
   \hspace{0.05\columnwidth}
   \includegraphics[width=0.33\columnwidth]{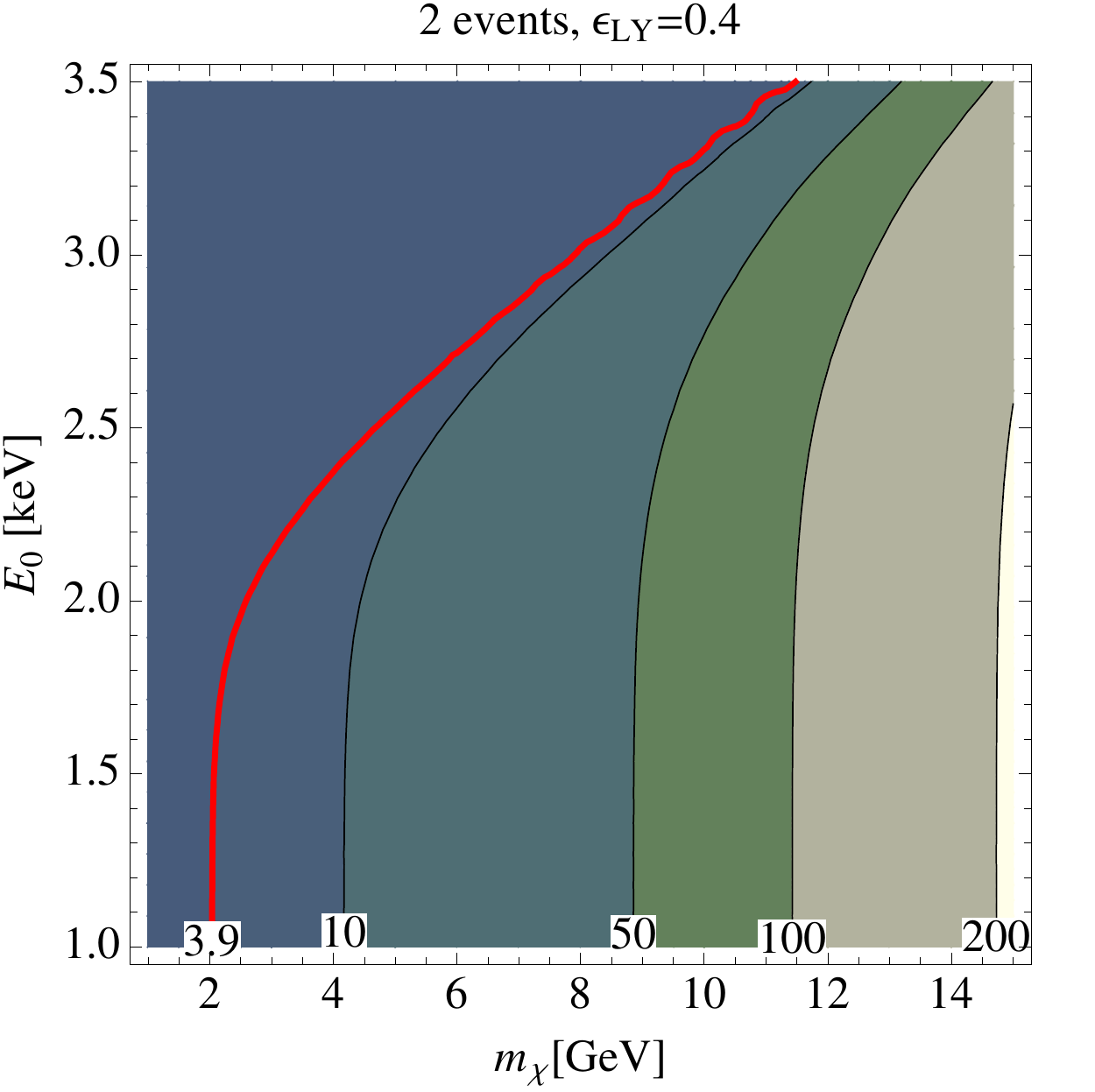} 
   \\
   \includegraphics[width=0.33\columnwidth]{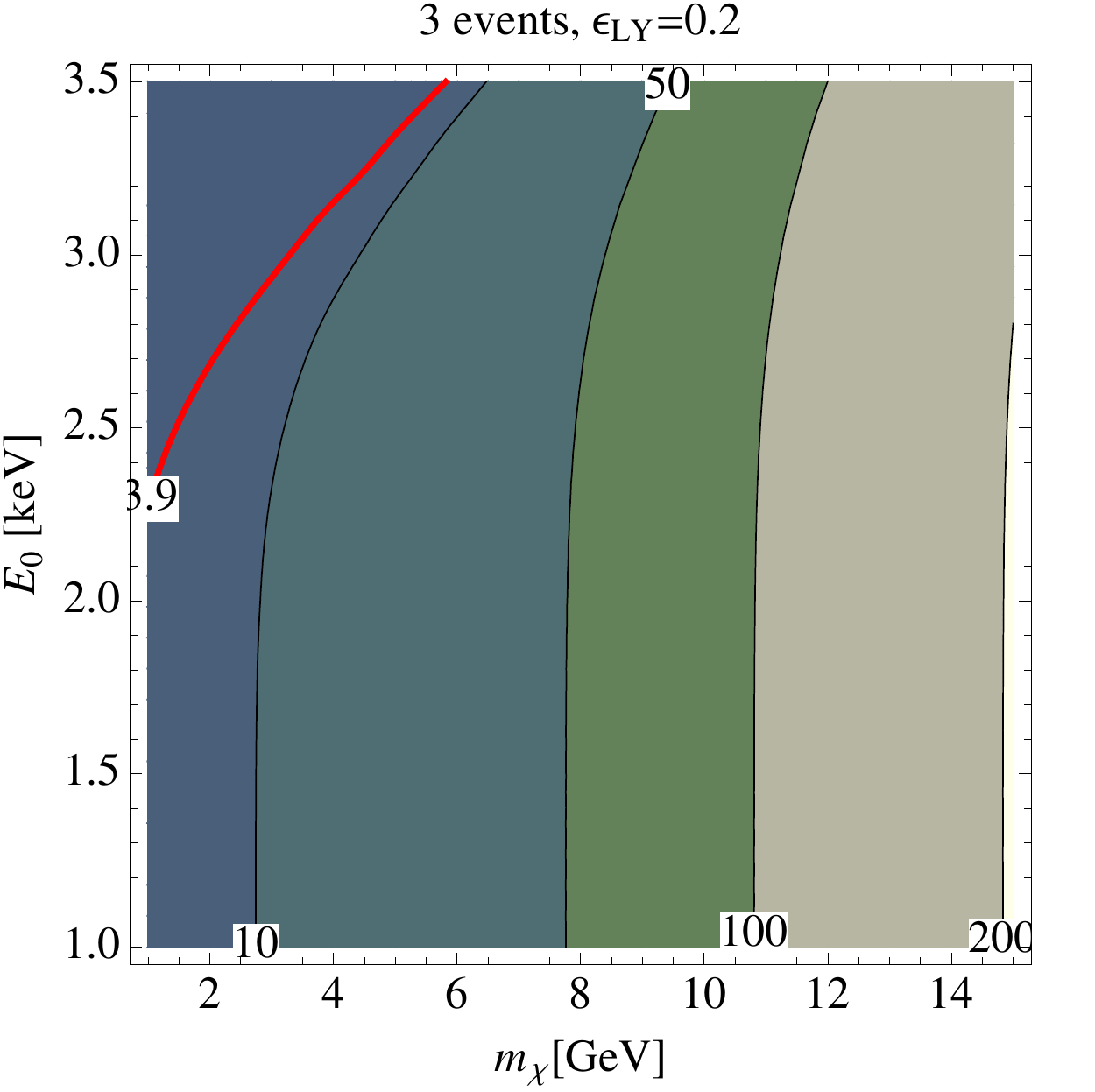}   
   \hspace{0.05\columnwidth}
   \includegraphics[width=0.33\columnwidth]{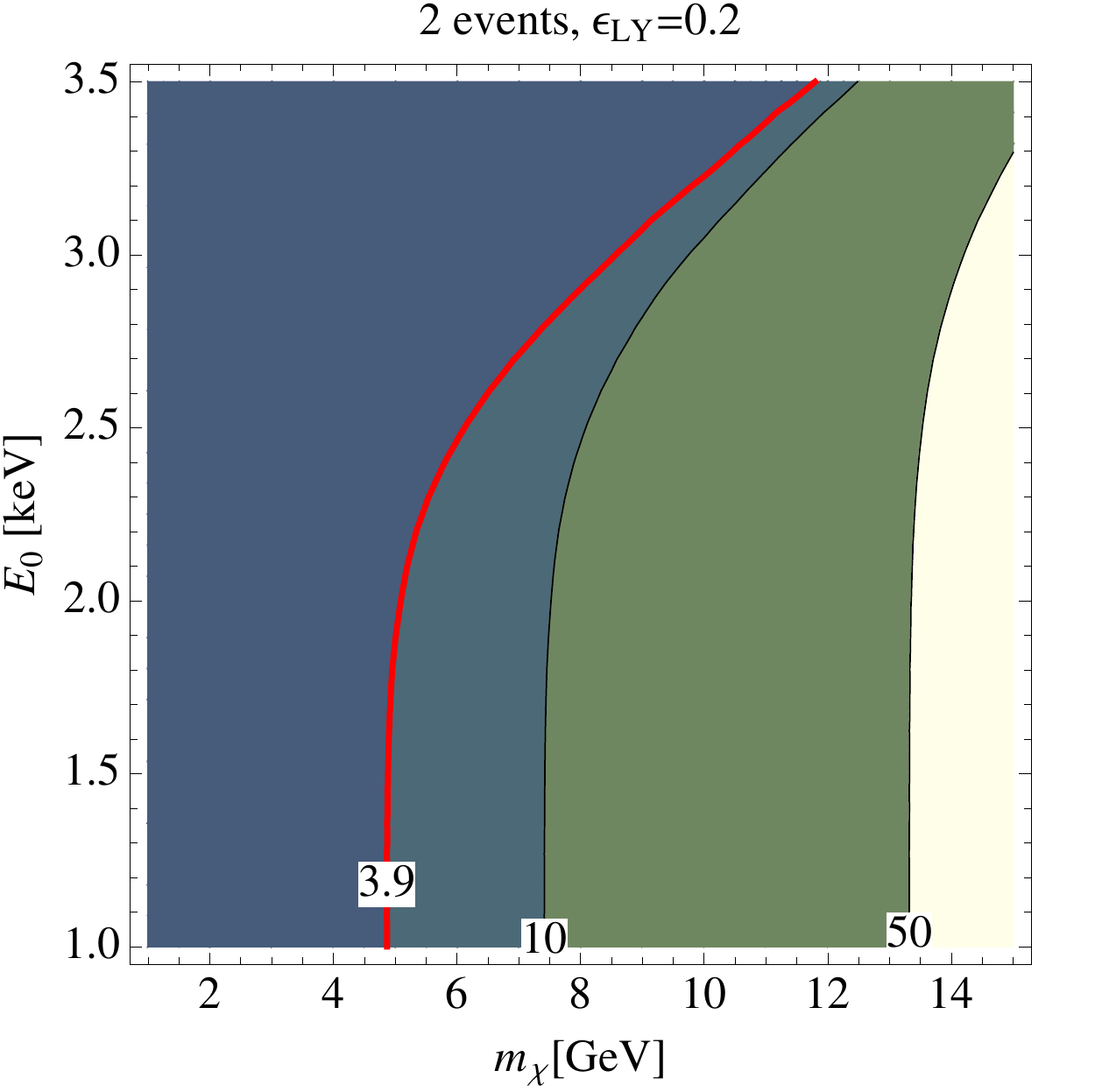} 
 \caption{The \emph{minimum} event rate at LUX, independent of astrophysics, for an efficiency as shown in Figure \ref{fig:LUXeffs} times a step function, $\Theta(E_R-E_0)$.  The rate is shown for various assumptions about the effect of the voltage on the light yield ($\epsilon_{\mathrm{LY}}$).  In all cases we consider the observation at CDMS a 10\% likely upward fluctuation, and we have taken $f_p=1$, $f_n=0$.}
   \label{fig:LUXpredictioneDM}
\end{figure}

Using the results of section~\ref{sec:astroindependent} we can determine, independent of astrophysics assumptions, the number of expected events at LUX.  We take the efficiency corrected rate to be flat below the highest event considered, Figure~\ref{fig:CDMSevents}  We consider the possibilities of all three events being due to DM, or just the lowest two.  As discussed earlier we also take the conservative assumption that the number of observed events at CDMS-Si was a (10\% likely) upward fluctuation on a smaller expected rate.  It is this rate that we map to LUX, while we set $f_p=1$ and $f_n=0$.  In section \ref{sec:IDDM} we consider the possibility of cancelation between the proton and neutron couplings to suppress the rate at LUX.  Although we believe our LUX efficiencies Figure~\ref{fig:LUXeffs} are accurate down to nuclear recoil energies of at least 2 keV we consider the possibility that they are an overestimate at low recoil energies and set the efficiency to 0, below some threshold $E_0$, i.e. $\epsilon(E_R)\rightarrow \Theta(E_R-E_0)\epsilon(E_R)$.  Furthermore, since the effects of the background electric field are not completely understood we investigate how varying the reduction in light yield affects the bounds.  We ignore the effects of energy resolution \cite{Frandsen:2011gi,Gondolo:2012rs}, although we have checked that its inclusion does not appreciably alter the result.
Under these rather conservative assumptions the \emph{minimum} event rate expected at LUX is shown in Figure \ref{fig:LUXpredictioneDM}.

The results encapsulated in Figure~\ref{fig:LUXpredictioneDM} can be qualitatively understood by returning to Figure~\ref{fig:vminplot}, which shows the model independent mapping of the energies of the three CDMS-Si events to recoil energies in xenon.  From this one can see the importance of both a possible 3 keV threshold and the energy of the third event at CDMS.  If one believes that the scintillation light from nuclear recoils below 3 keV is zero, then the two lower energy events will not produce a signal at LUX for DM lighter than $\sim 8$ GeV. However, we should stress that the insensitivity of LXe to the CDMS signal is arising from two simultaneously very conservative assumptions: that no light is produced for these lower energy events, and that there are essentially no particles with higher velocity - an assumption that is highly unlikely unless the WIMP is lighter than $\sim$  GeV. If either of these assumptions is relaxed, then we would expect sensitivity from the LXe experiments.

From Figure \ref{fig:LUXpredictioneDM} we see that if one wishes to believe that all 3 events at CDMS are due to DM scatters and our LUX efficiencies, with the nominal $\epsilon_{\mathrm{LY}}=0.8$, are accurate above 3 keV then DM must be lighter than $\sim 2$ GeV to avoid the LUX bounds.  Note, however that for such light DM the 12.3 keV event at CDMS-Si would require a $\vmin$ of 1600 km/s, well above the galactic escape speed in the Earth's frame, Figure~\ref{fig:vminplot}.  This tension is not greatly reduced by lowering the light yield, for $\epsilon_{\mathrm{LY}}=0.2$ the mass bound is raised to $\sim 3$ GeV, with a correspond $\vmin$ of 1400 km/s.  If, however, the highest event at CDMS-Si is due to background and the LUX efficiency is only non-zero above 3 keV the allowed parameter space is more reasonable with masses of $\sim 8$ GeV corresponding to $\sim 600$ km/s speeds.  Again lowering the light yield does not markedly change the result unless lower recoil energies, below 3 keV, where the efficiencies become over an order of magnitude different, are being probed.

Note that this event rate at LUX is the \emph{minimum} allowed since it makes no assumptions about the velocity distribution, other than it is positive and thus $g(\vmin)$ is a monotonically decreasing function.  For the conventional case of a MB distribution the tension is even larger since the velocity distribution grows as an exponential at low speeds.  For instance, the best fit point for $f_p=1$, $f_n=0$ ($m_\chi=8.6$ GeV, $\sigma = 7.6\times 10^{-41}$ cm$^2$) predicts $\sim 980$ events, above 3 keV recoil energy, at LUX for the nominal case of $\epsilon_{\mathrm{LY}}=0.8$.  This is lowered to $\sim 170$ if $\epsilon_{\mathrm{LY}}=0.2$.  Our result confirms the tension observed for a MB velocity distribution between the CDMS results and the bound from LUX, and emphasises that this tension cannot be relieved by altering astrophysics alone, instead one must make further assumptions about detector behavior and the number of signal events at CDMS-Si.

Clearly, even for these \emph{very} conservative set of assumptions, there is considerable tension between CDMS and LUX.  The results can be put into some level of agreement only if some of the CDMS events are due to background, or LUX has overestimated its efficiency.

\subsection{Isospin dependent Dark Matter}\label{sec:IDDM}

The scattering rate of DM off the target material in a direct detection experiment depends on the isotopic content of the detector, (\ref{eq:mastereq}).  For detectors made up of various elements, or isotopes of the same element, with atomic numbers and mass number $(Z_i,A_i)$ each with mass fraction $\kappa_i$, the rate for any individual element scales as,
\be
C_T^{(i)}=\kappa_i (f_p Z_i + f_n(A_i-Z_i))^2~.
\ee
Thus, for any isotope there is a particular value of $f_n/f_p$ for which $C_T=0$.  This is a leading order statement, and we discuss corrections to this below.  In an experiment where a single isotope overwhelmingly dominates (silicon has one isotope with abundance over 90\%) it is possible to essentially make that experiment blind to DM.  For LUX there are five isotopes that make up over 90\% of the mass of the target so, although their mass numbers are within 1\% of each other, the cancelation will not be complete but can still help remove a large fraction of the tension between CDMS and LUX.  The total rate at LUX scales as $\sum C_T^{(i)}$, and the minimum rate occurs if $f_n\approx -0.7 f_p$.  For isospin dependent models (IDDM)\cite{Feng:2011vu}, $f_n\ne f_p$, that lower the rate, this cancellation occurs to varying extents for all elements.  So if this is the explanation of the CDMS events the intrinsic quark-DM coupling must be larger than in models without this cancellation, making collider constraints \cite{Beltran:2010ww,Goodman:2010yf,Goodman:2010ku,Rajaraman:2011wf,Fox:2011pm,Shoemaker:2011vi,Fortin:2011hv,Fox:2011pm,Fortin:2011hv,Bai:2012xg,Fox:2012ee} a probe of these models\cite{Feng:2013vod}.

\begin{figure}[t] 
   \centering
    \includegraphics[width=0.33\columnwidth]{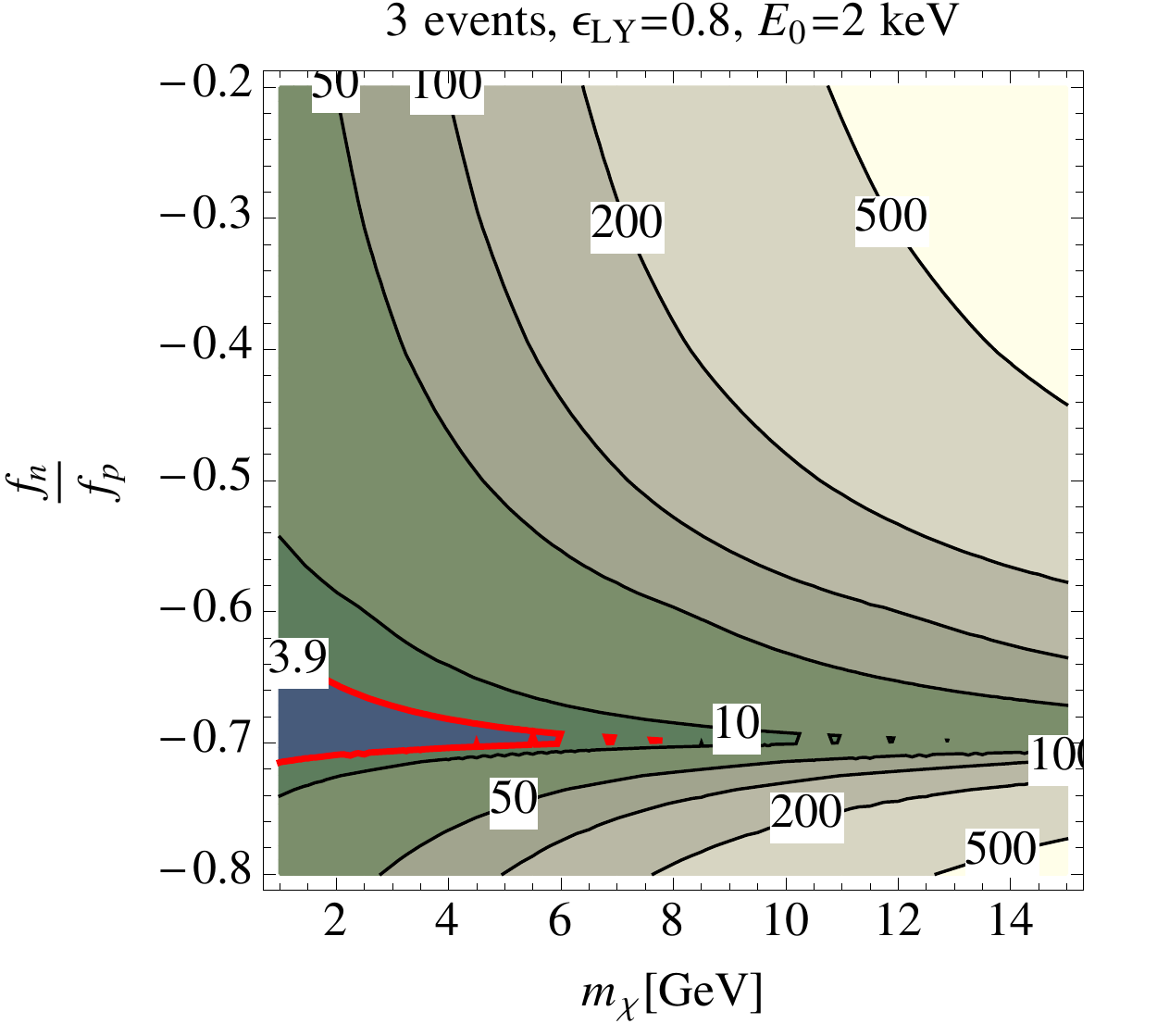}
     \includegraphics[width=0.33\columnwidth]{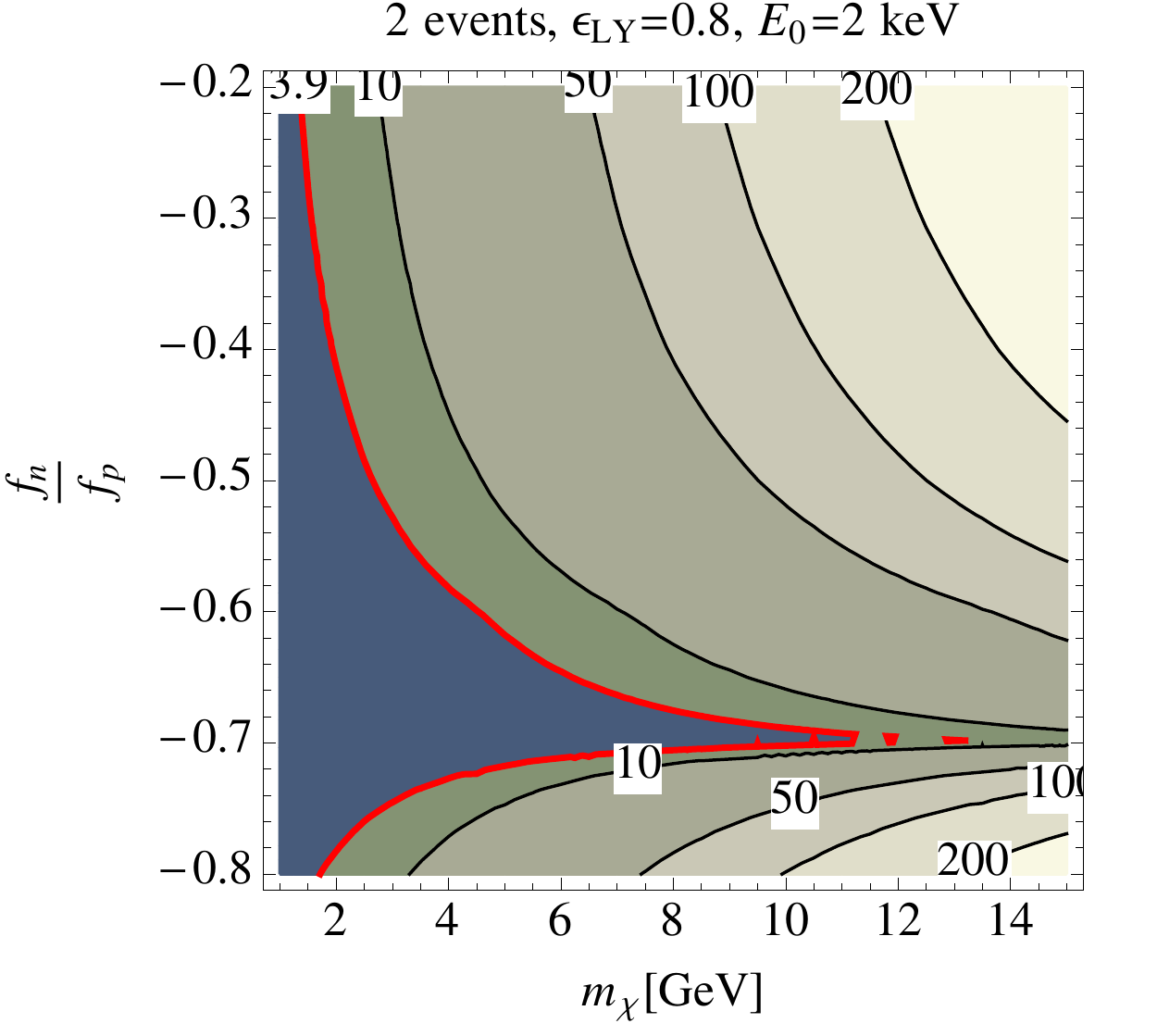}
     \\
     \includegraphics[width=0.33\columnwidth]{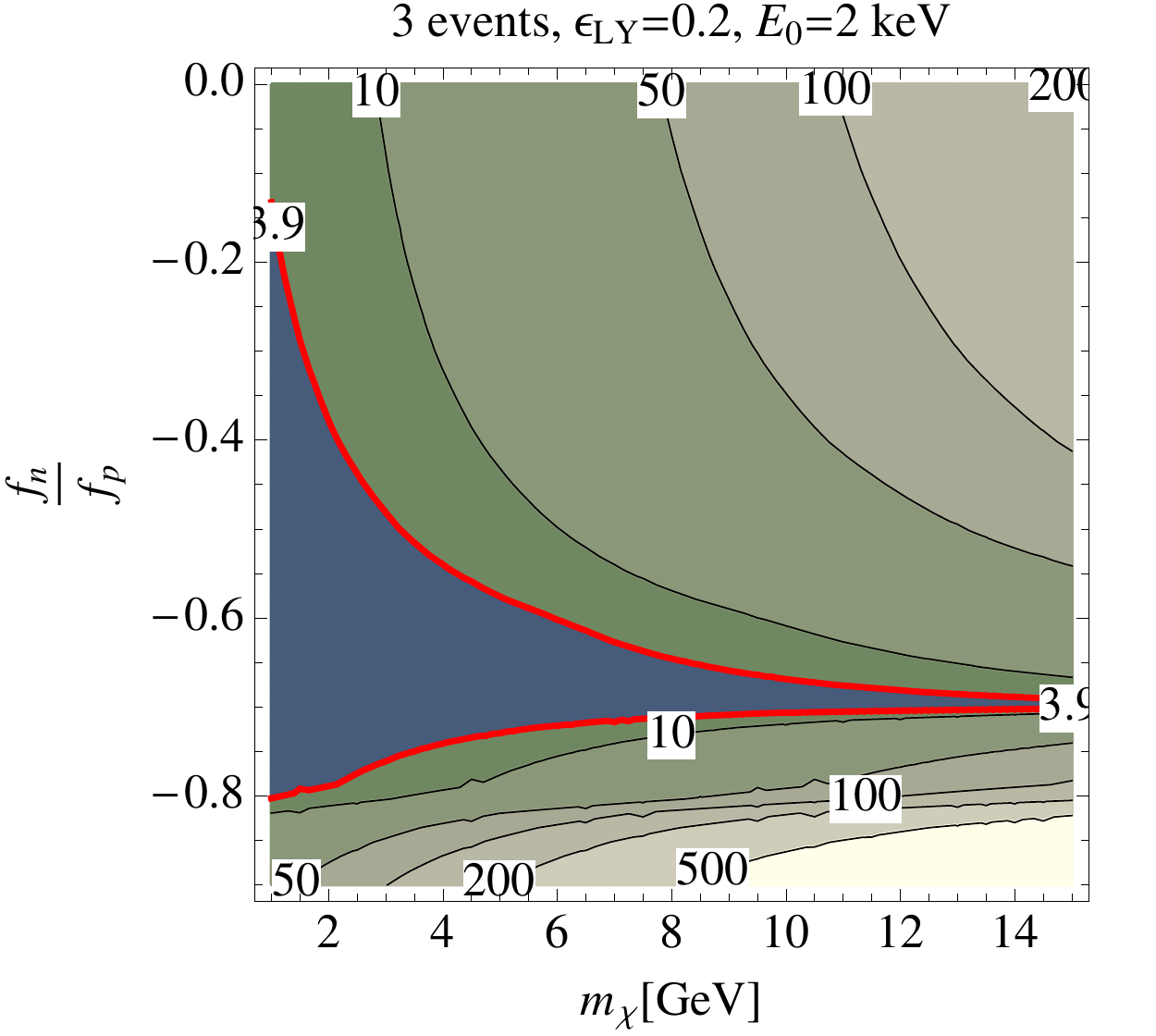}
     \includegraphics[width=0.33\columnwidth]{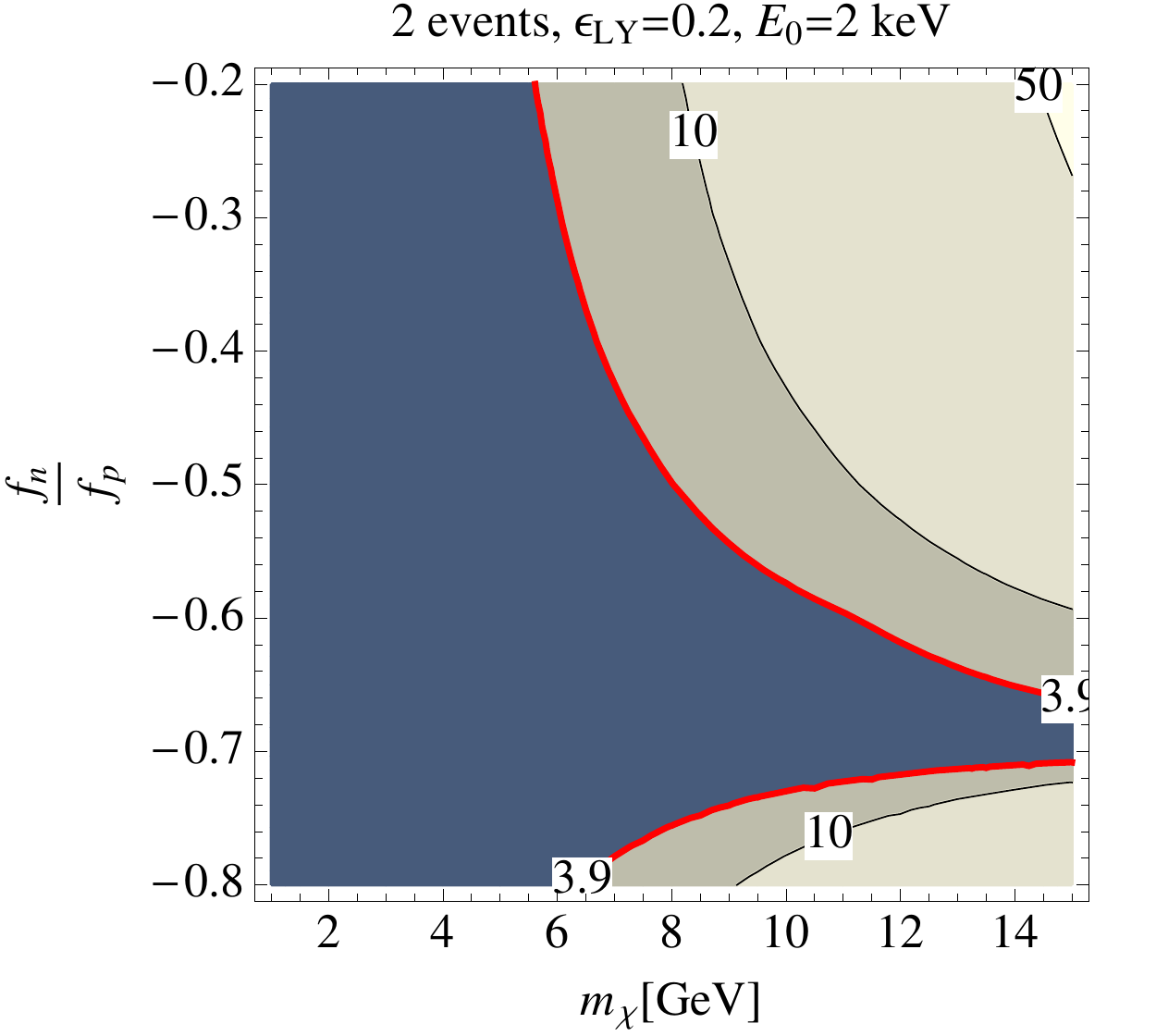}
     \\
     \includegraphics[width=0.33\columnwidth]{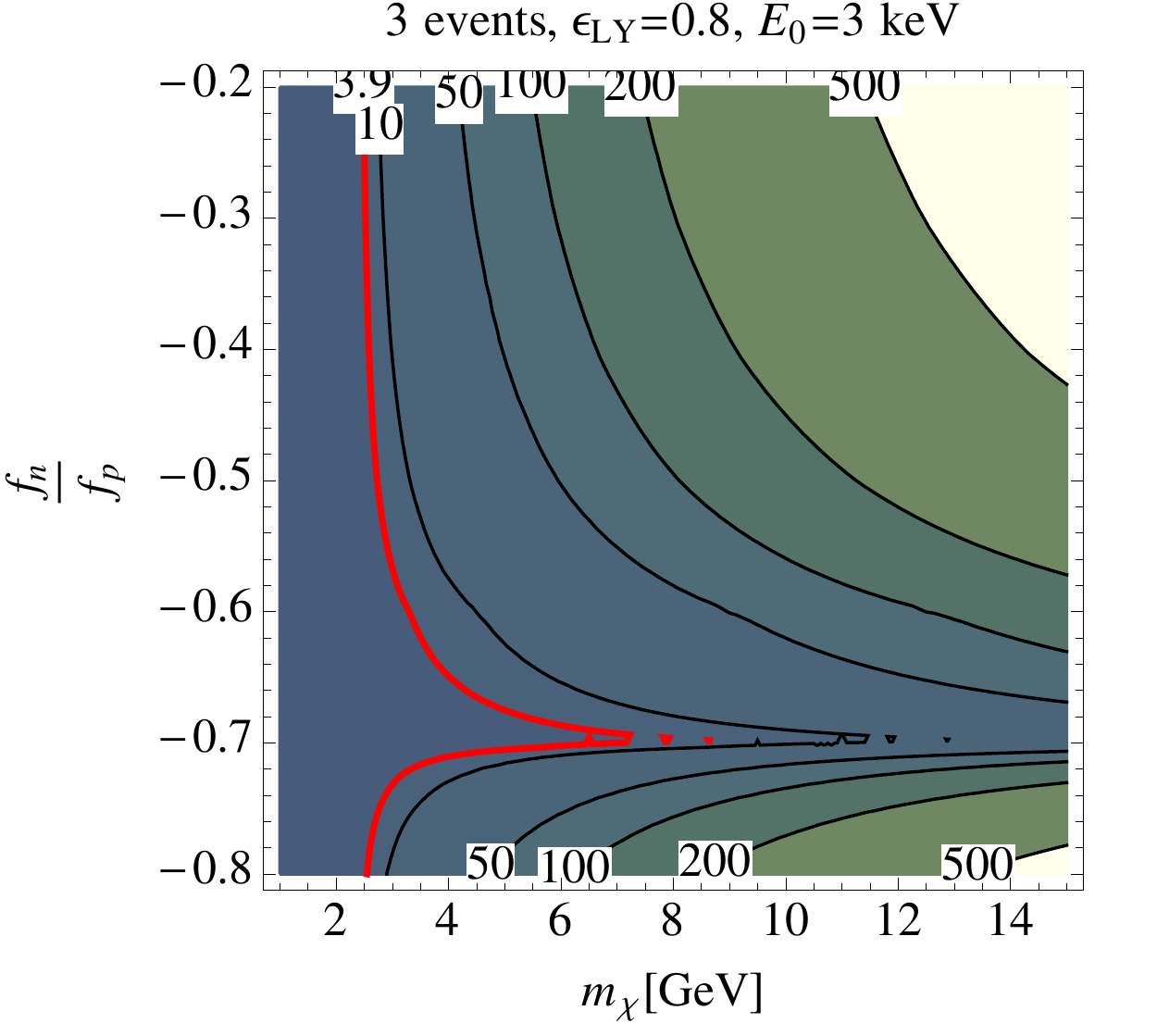}
     \includegraphics[width=0.33\columnwidth]{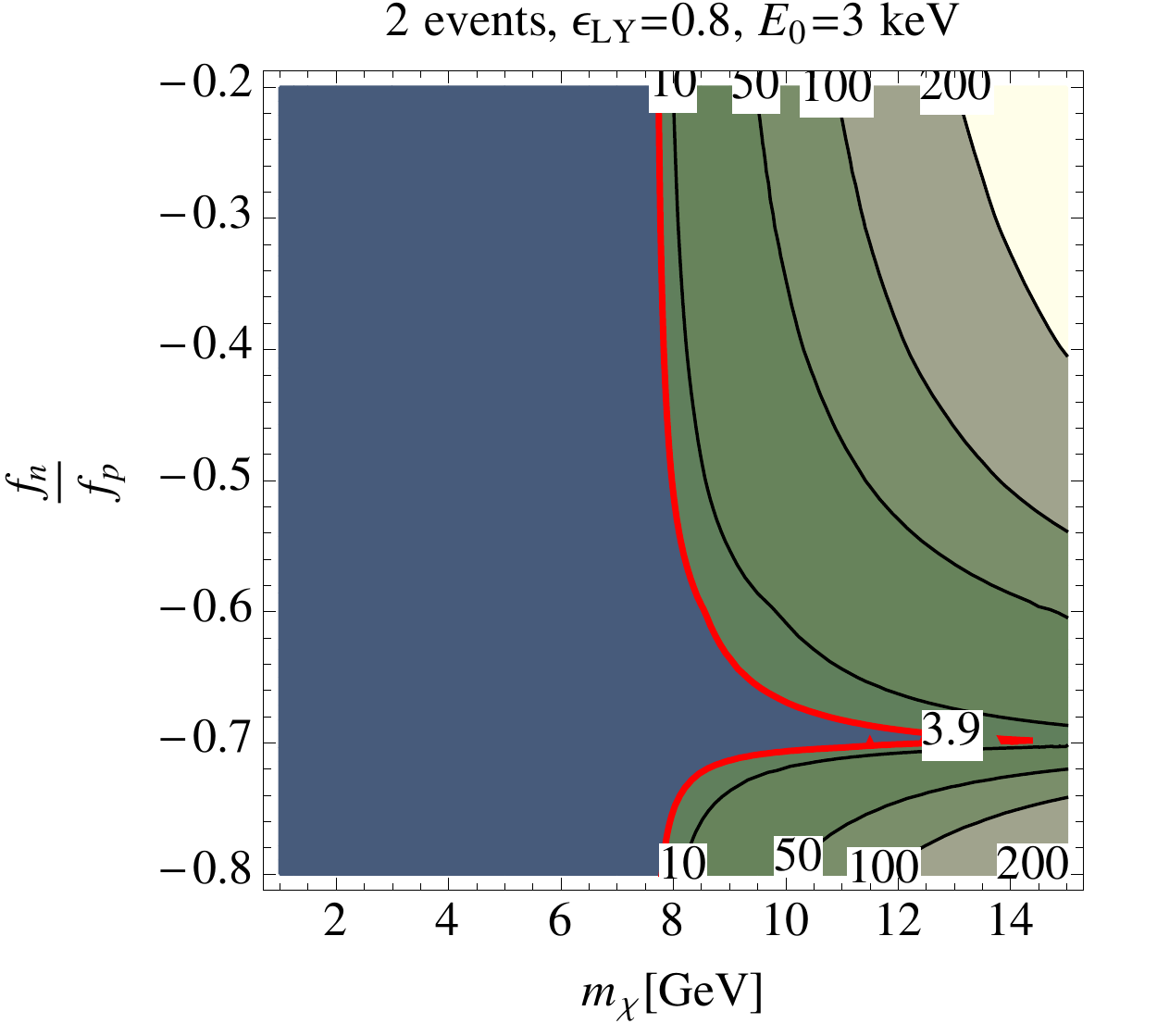}
   \caption{The expected number of events at LUX if DM couples in an isospin dependent fashion $(f_n\ne f_p$).  The right-hand plots assume the highest energy event at CDMS is not due to DM, whereas the left-hand plots consider all three as signal.  The upper plots use the nominal efficiency function for LUX, the middle plots take $\epsilon_{\mathrm{LY}}=0.2$, and the lower plots assume the threshold is 3 keV.  The efficiency functions are shown in Figure~\ref{fig:LUXeffs}.}
   \label{fig:isospindependent}
\end{figure}

For $f_n\approx -0.7 f_p$ the rate at LUX will be suppressed but the cancellation is not complete because of the many stable isotopes of xenon that are in the experiment.  Compared to the case of $f_n=0$, where the ratio\footnote{We include all isotopes present in CDMS-Si and LUX.} $C_T(Xe)/C_T(Si)$ is $\sim 15$, the isospin dependent ratio is over 100 times smaller.  This allows the two results to become compatible, however, as can be seen in Figure \ref{fig:isospindependent}, there is a very small range over which the rate at LUX can be made small enough.  If the detector works as expected down to 2 keV and all three CDMS events are signal the DM must also be very light and as described above the necessary speeds are above expected escape velocities.  Isospin dependent DM only seems a viable way of relieving the CDMS-LUX tension if further assumptions are made: not all CDMS events are signal, the light yield is considerably lower than expected, or the detector is insensitive to recoils below 3 keV.

The large cancellation between the neutron and proton form factors that allows the rate at LUX to be sufficiently suppressed assumes that the form factors are independent of energy.  However, as has recently been demonstrated for the case of scalar couplings between DM and quarks~\cite{Cirigliano:2012pq}, one loop corrections and two-nucleon amplitudes alter these form factors making them momentum dependent. Thus, this cancellation is no longer achievable at all recoil energies.  Although these higher order corrections are not enhanced beyond what is naively expected, they can become very important when the leading contribution is suppressed, as in IDDM~\cite{Cirigliano:2012pq,Cirigliano:2013zta}.

Given the discussion above, it is interesting to see if it also can explain other positive results, such as the modulation seen at DAMA\cite{Bernabei:2010mq} in the region of parameter space for which IDDM makes CDMS and LUX compatible.  Interestingly, the events at CDMS occurred in the half of the year where one would expect the rate to be highest - the events were spaced over March to September of the same year.  

For the light DM necessary to explain CDMS the scattering at DAMA will be off sodium.  The cancellation that suppresses the rate at LUX will also suppress the rate at DAMA.  Since sodium is lighter than silicon, the rate at DAMA is lower even before the IDDM cancelation.  The suppression for scattering off sodium is only about 35\%, the ratio $C_T(\mathrm{Na})/C_T(\mathrm{Si})\approx0.4$ if $f_n\approx -0.7 f_p$ compared to $\approx 0.6$ if $f_n=0,\,f_p=1$.    Since CDMS only saw events between March and September, it is possible that the rate is almost 100\% modulated, that is there would be 0 events in the winter, and only events around the peak date of $\sim$ June 2$^{\mathrm{nd}}$.  For this maximal modulation, and taking all three observed events as a (10\% likely) downward fluctuation on a true signal of 6.7, the size of the modulation amplitude one would expect at DAMA is 
comparable to the observed modulation rate of $\sim0.02$/cpd/kg/keV.  As can be seen from Figure~\ref{fig:isospindependent} there is room to move away from maximal suppression and still avoid present LUX constraints.

\subsection{Exothermic DM}

The final class of models we wish to consider are those involving down scattering of the DM, so called exothermic DM   \cite{Graham:2010ca}.  Unlike up scattering models, inelastic DM \cite{TuckerSmith:2001hy}, which tend to favor detectors built from heavy elements, light exothermic DM is best searched for using light elements.  Thus, we investigate the possibility that the excess at CDMS-Si is due to exothermic DM and determine if this can be made consistent with the bounds from LUX.  

In exothermic DM the incoming dark state is heavier than the outgoing dark state by $\delta$.  In order that the heavier state is sufficiently long lived that there are enough of them in our vicinity to give a signal in direct detection experiments this splitting cannot be too large.  We will consider $|\delta|\ltap 100$ keV.  For such small splittings the only available decays are to neutrinos or photons.  If the couplings to the SM occur through kinetic mixing of a dark sector gauge boson with the SM gauge bosons the lifetimes are longer than the age of the universe \cite{Finkbeiner:2009mi,Batell:2009vb}.

For negative $\delta$ the kinematics of the collision is dominated by the energy release (\ref{eq:vmin}), with the majority of the events having recoil energy $E_R=|\mu \delta/m_N|$.  The modulation fraction is very small and so this cannot explain the DAMA or CoGeNT modulation results.  Although the bulk of the events are offset from zero recoil, unlike eDM, the peak at $E_R=|\mu \delta/m_N|$ has a width set by the kinetic energy of the incoming DM.  If the peak falls close to the CDMS threshold there will be a non-negligible number of events below the threshold from DM colliding at non-zero speeds.  

\begin{figure}[t] 
   \centering
   \includegraphics[width=0.4\columnwidth]{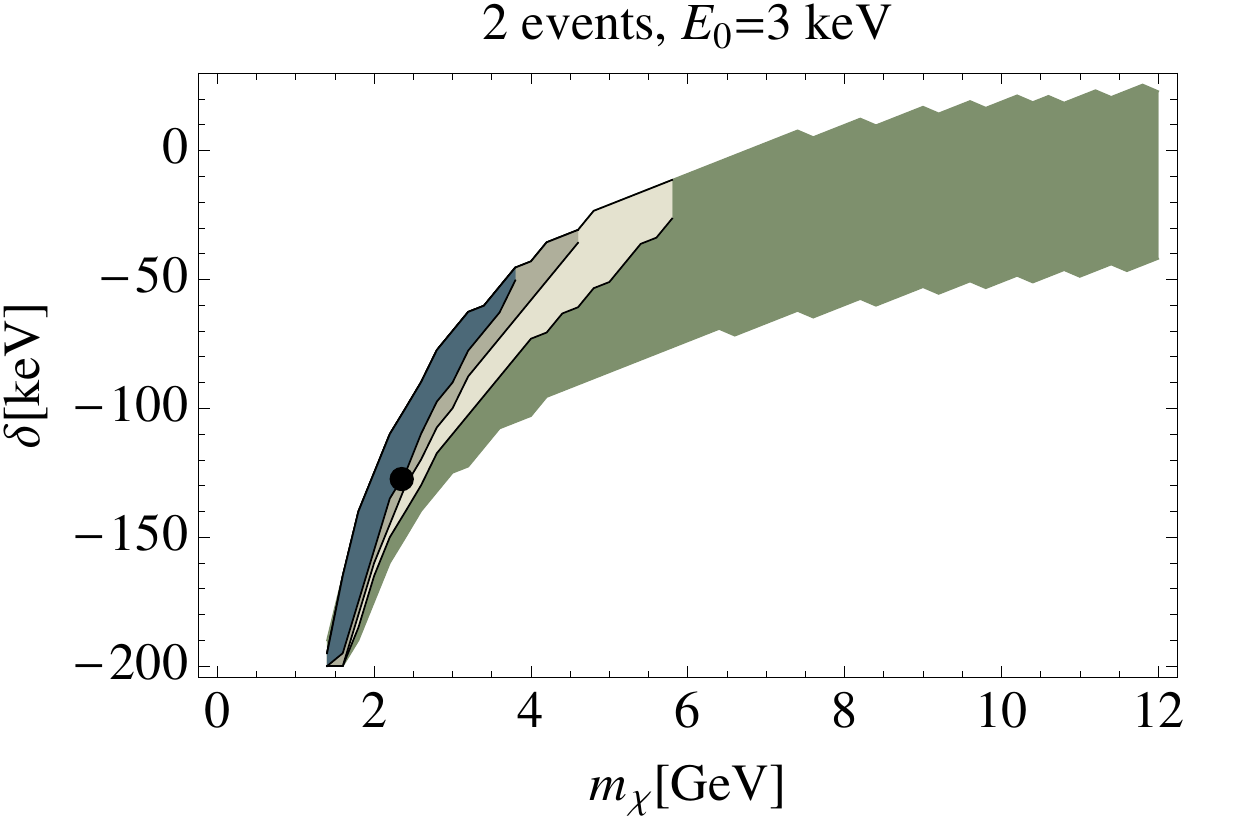} 
    \includegraphics[width=0.4\columnwidth]{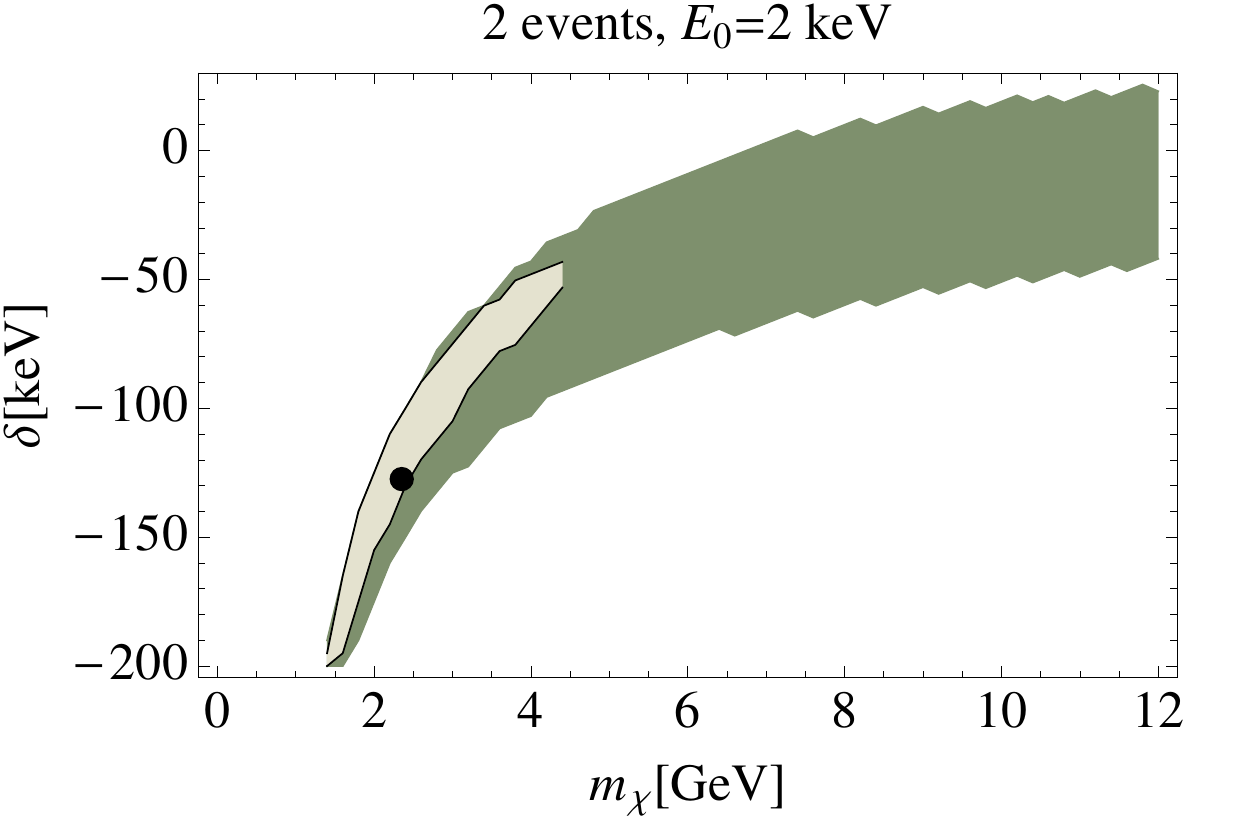}
    \\
    \includegraphics[width=0.4\columnwidth]{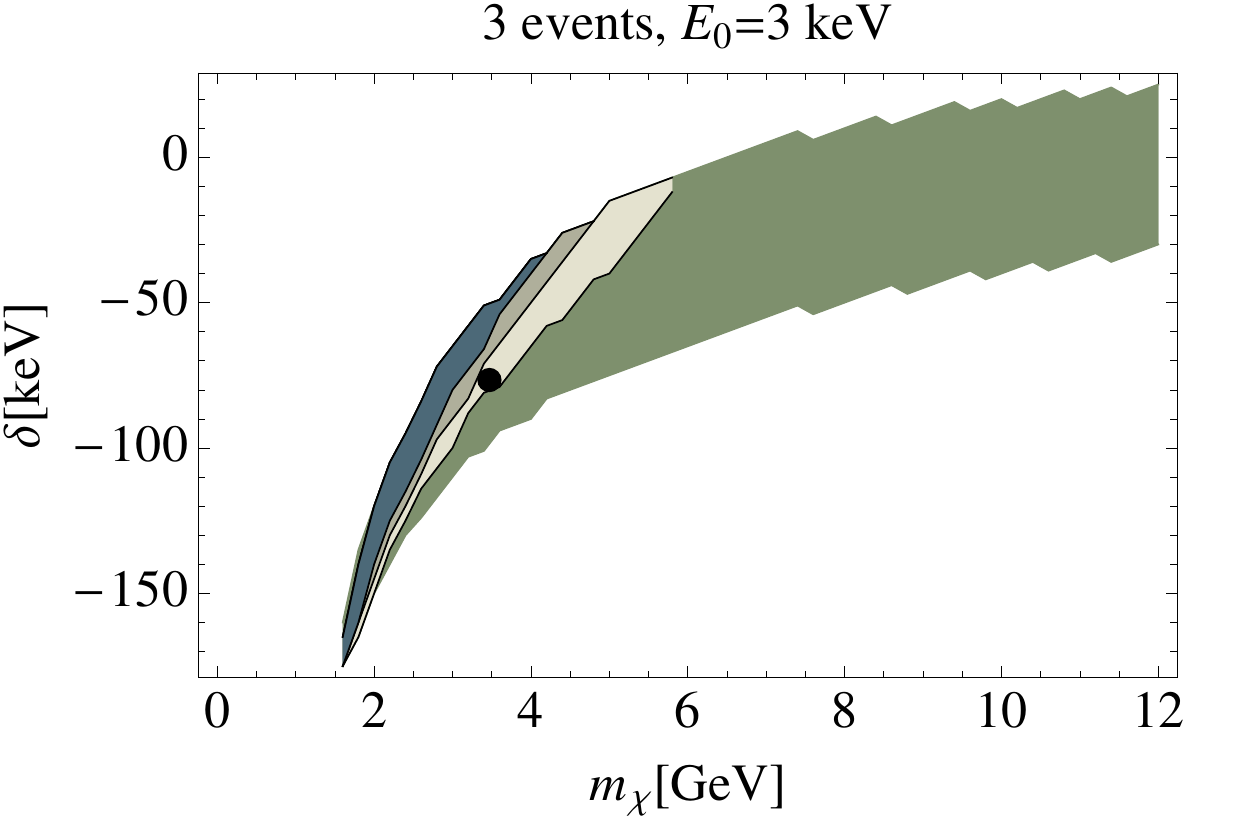} 
    \includegraphics[width=0.4\columnwidth]{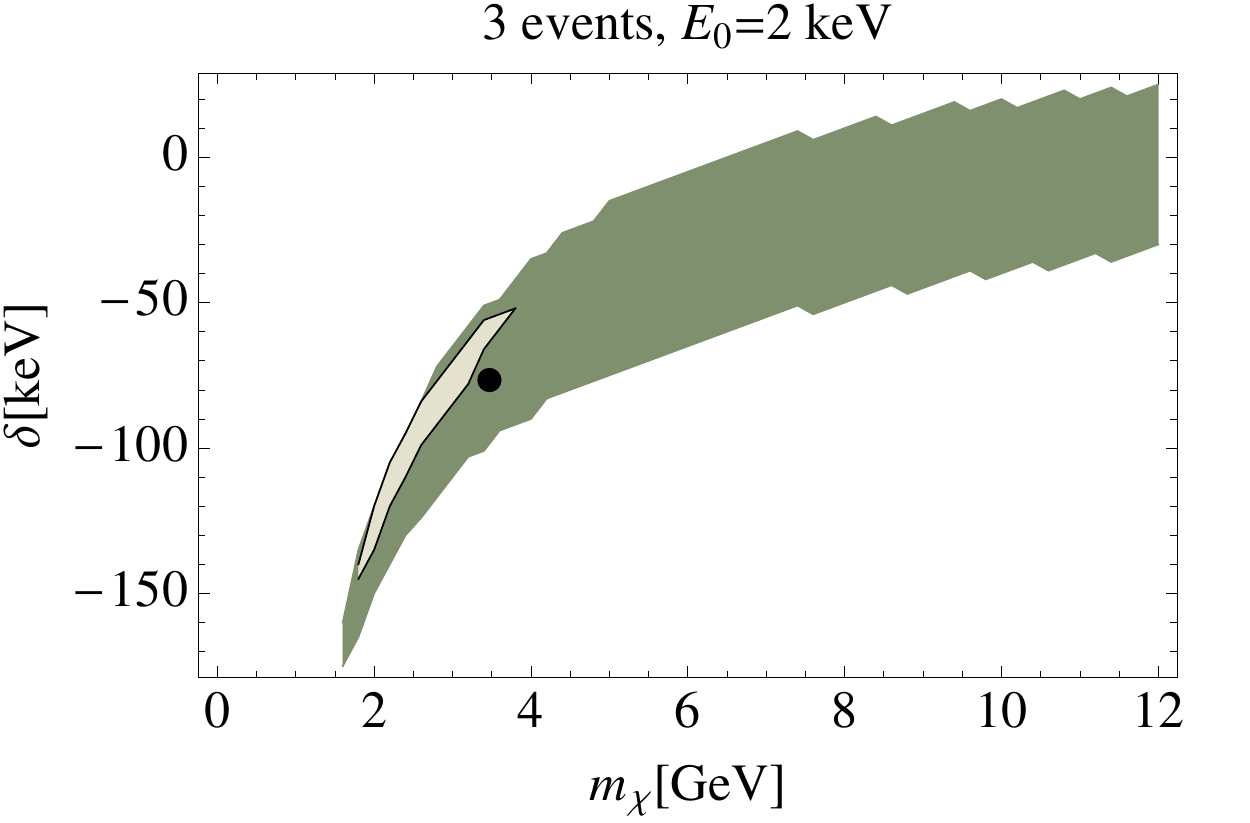}
   \caption{The event rate at LUX, assuming a Maxwell Boltzmann distribution and $f_n=0$, $f_p=1$.  The large green region is the region preferred by CDMS-Si, projected on the $m_\chi-\delta$ plane.  The overlapping smaller regions correspond to regions consistent with LUX assuming $\epsilon_{\mathrm{LY}}=0.8,0.4,0.2$, in reverse order of size.  The black dot denotes the best fit point to the CDMS data.\label{fig:exoLUX}}
\end{figure}

We start out by carrying out an analysis assuming the velocity distribution is a MB with parameters as given above.  We use the log-likelihood method to find the region of $(m_\chi,\sigma,\delta)$ space that, with 90\% confidence, explains the CDMS-Si result under the assumption that all three, or the two lowest energy, events are signal.  Using the efficiencies, as shown in Figure \ref{fig:LUXeffs}, for the three different light yields we consider, we then determine what part of this parameter space is consistent with the observations at LUX.  We determine the LUX bounds, again at 90\% C.L., using the max-gap method \cite{Yellin:2002xd}. This technique does not require that we know the form of the background but does take into account the energy of observed events, where we determined the most likely energy for the one event that falls in the signal region to be around 6 keV.  The technique also takes into account the range of energies observed and we take this to start at either 2 or 3 keV, the upper end does not affect the result once it is above $\sim 10$ keV.  The regions of parameter space, projected onto the $m_\chi-\sigma$ plane, that explain CDMS-Si and explain CDMS while being consistent with LUX are shown in Figure \ref{fig:exoLUX}.

Assuming $f_n=0$, $f_p=1$ the best fit point for the case of all three events being signal is $(m_\chi,\sigma,\delta)=(3.5\, \gev,\, 1.2\times 10^{-42}\,\mathrm{cm}^2,\, -77\,\kev)$, and if the highest energy event is background the best fit is $(2.4\,\gev,\, 5\times 10^{-43}\mathrm\,{cm}^2,\, -127\,\kev)$.  It is clear from Figure \ref{fig:exoLUX} that in all cases there is only a small region of the parameter space that is consistent with CDMS-Si and LUX, at 90\% C.L..  If our estimate of the LUX efficiency is accurate down to 2 keV, as we expect, this region of parameter space is further shrunk.

\subsubsection{Exothermic Astroindependent}

The astrophysics independent comparison of non-elastic DM is more involved, but still tractable and has recently been discussed \cite{meKITP,Bozorgnia:2013hsa,DelNobile:2013cva}.  Here we apply the astrophysics independent approach as outlined in section \ref{sec:astroindependent} to the case of exothermic DM.

To quantify the amount, or lack, of consistency between the two results we calculate the predicted $\tilde{g}(\vmin)$ from the CDMS-Si results, assuming all events are due to signal, as well as the 90\% C.L. on the results, by assuming each bin was a fluctuation on the true rate.  We bin the data in bins of 2 keV and in mapping from energy space to $\vmin$-space the bin widths are determined by finding the minimum and maximum $\vmin$'s within each energy bin.  

\begin{figure}[t] 
   \centering
 \includegraphics[width=0.4\columnwidth]{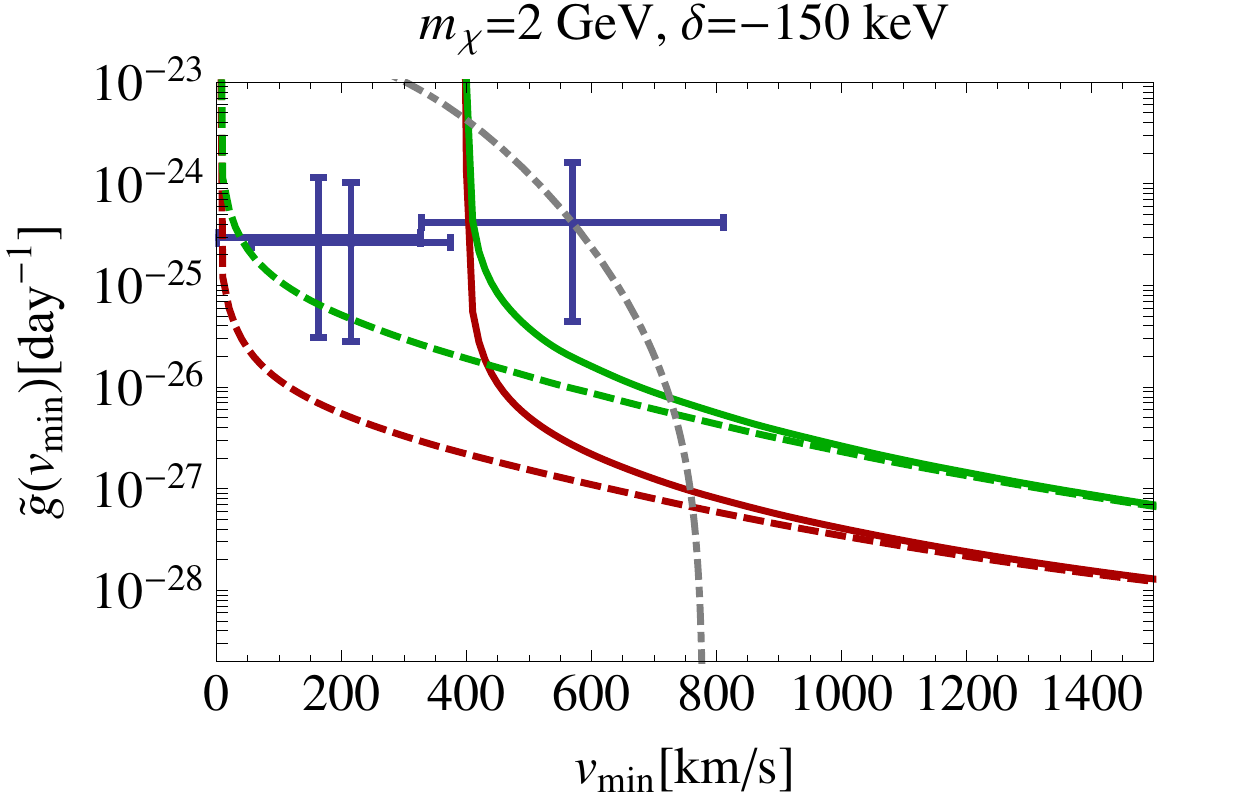} 
\includegraphics[width=0.4\columnwidth]{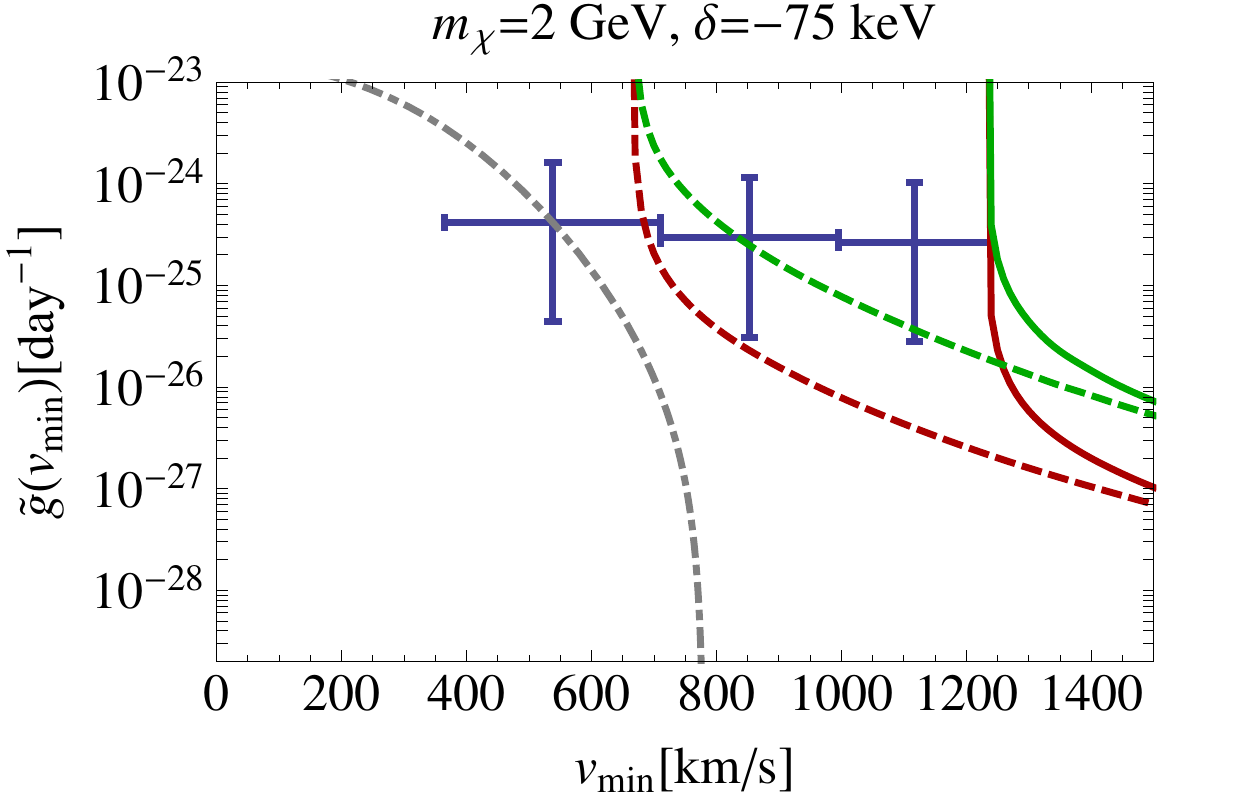} 
\\
\includegraphics[width=0.4\columnwidth]{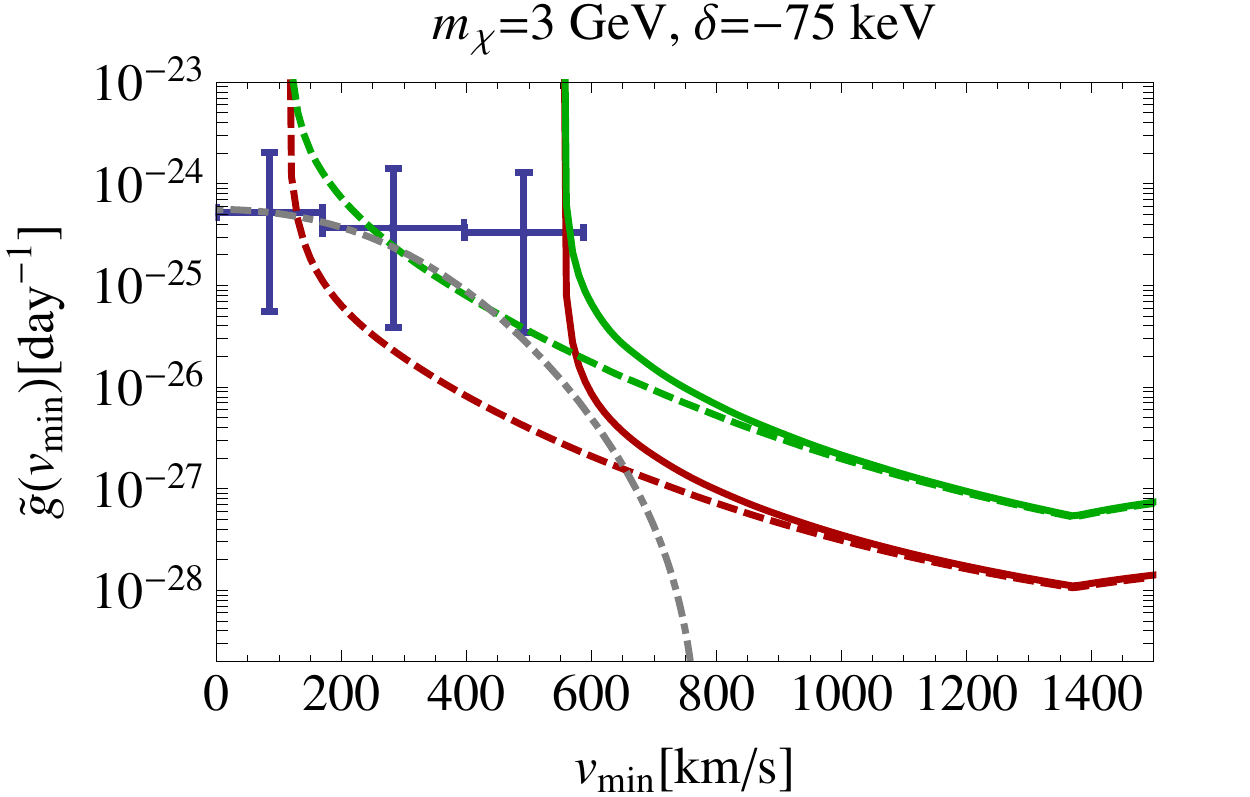} 
\includegraphics[width=0.4\columnwidth]{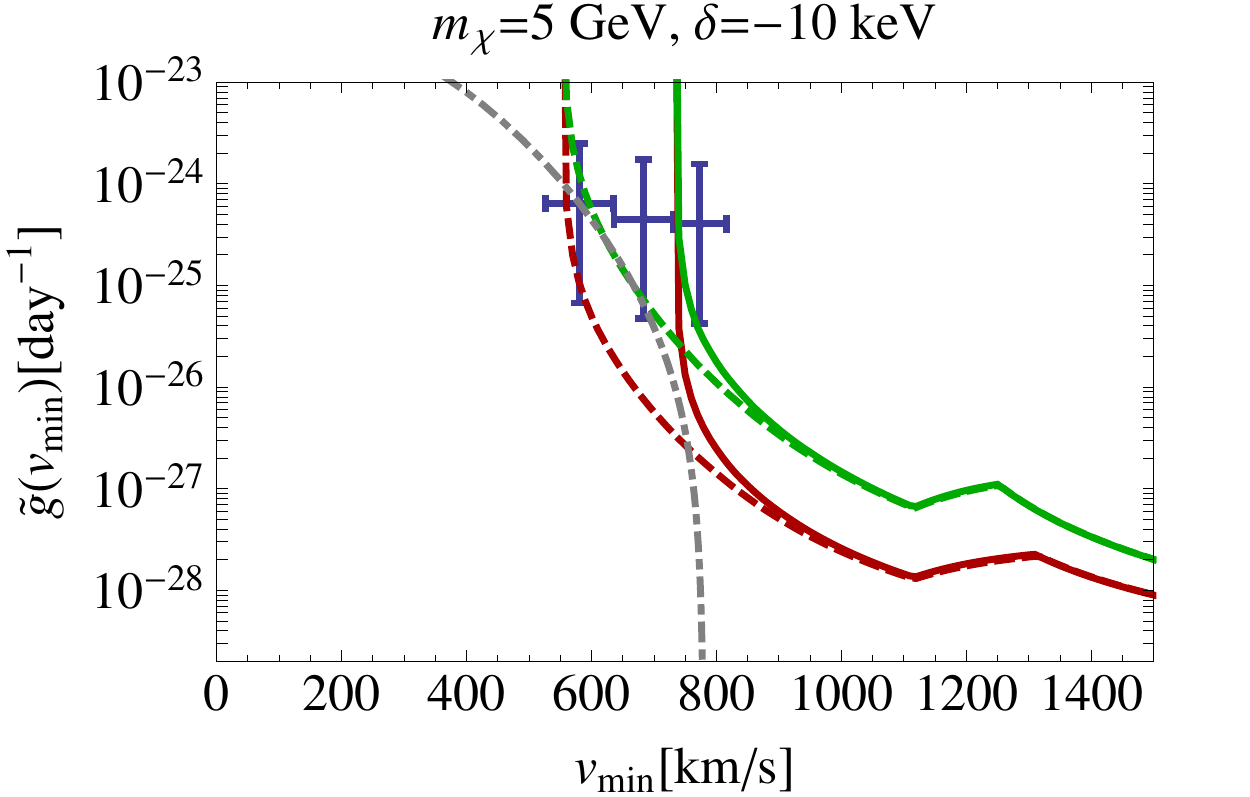} 
\\
\includegraphics[width=0.4\columnwidth]{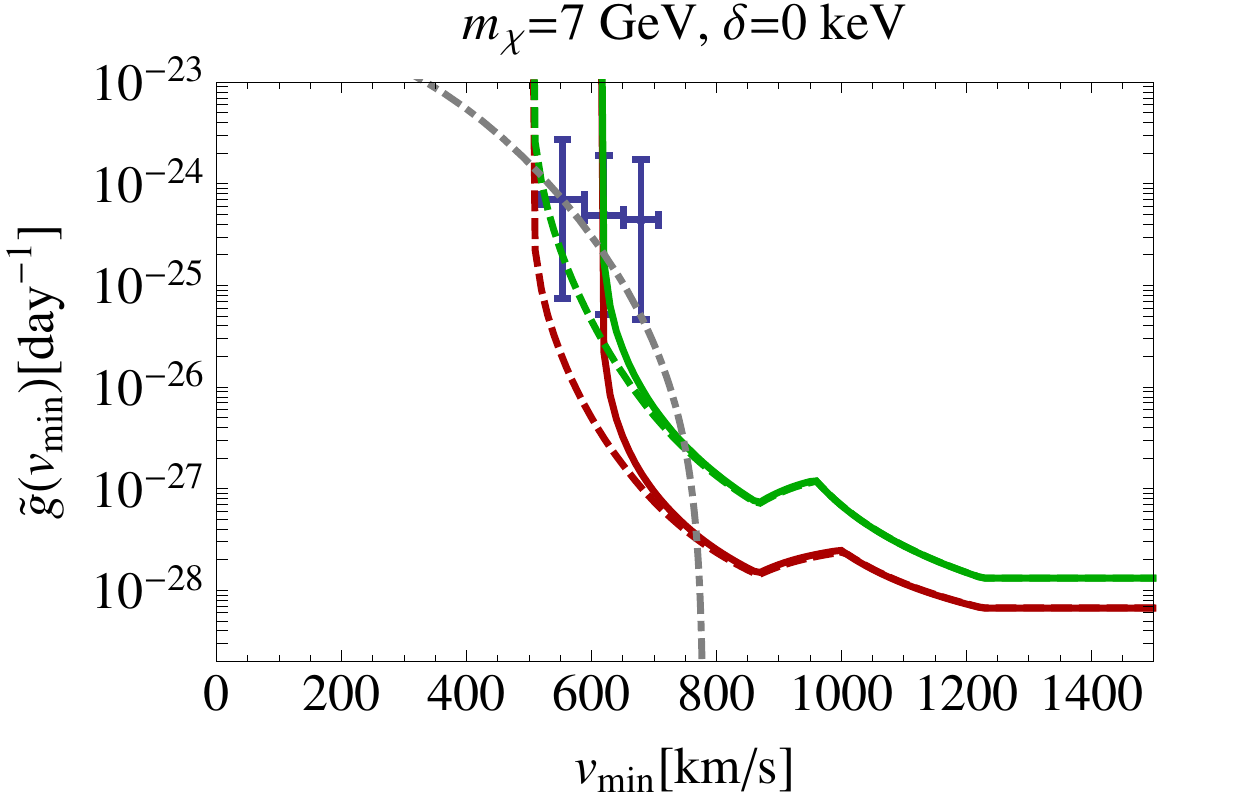} 
\includegraphics[width=0.4\columnwidth]{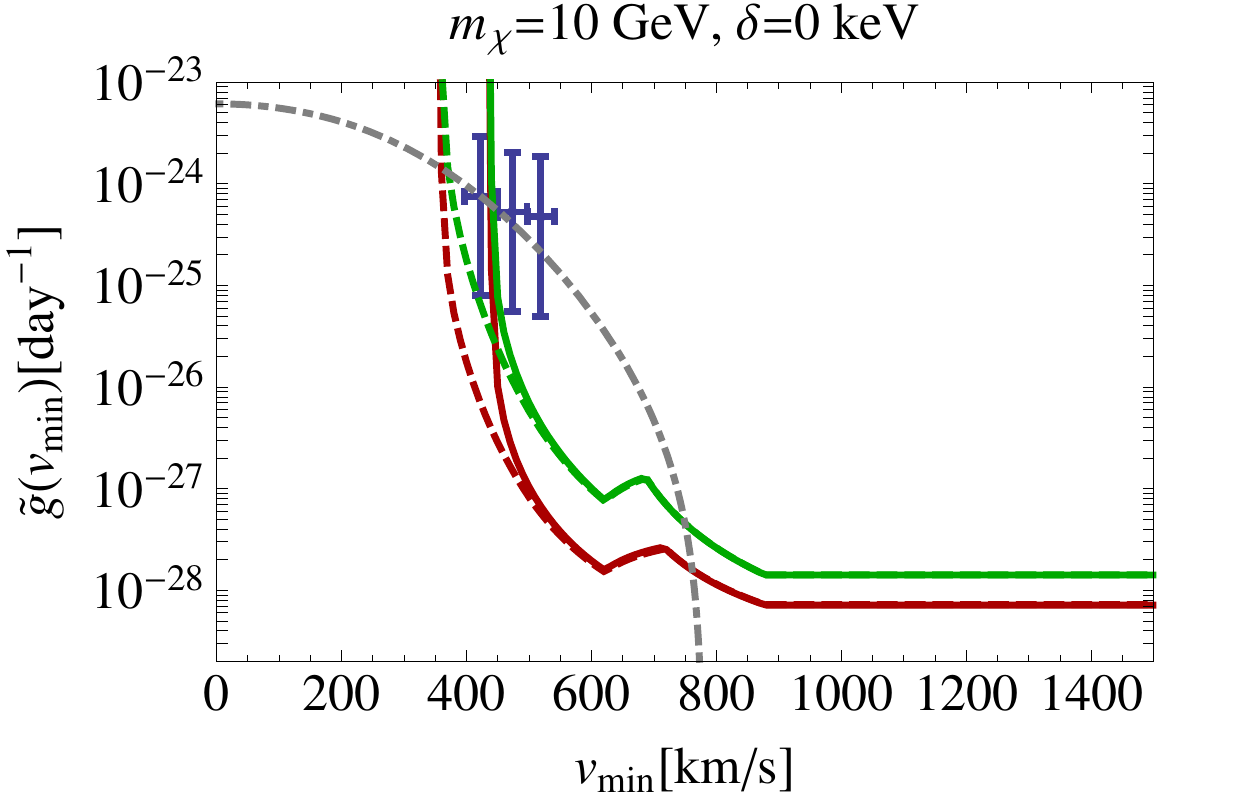} 
\caption{Comparing the prediction for $\tilde{g}$ as determined by CDMS-Si to the constraint from LUX for particular parameters.  The solid curves are for a threshold of 3 keV and the dashed for 2 keV, the red curves correspond to $\epsilon_{\mathrm{LY}}=0.8$ and the green to $\epsilon_{\mathrm{LY}}=0.2$, the grey dot-dashed curve is the MB prediction, normalised to agree with the prediction from the lowest energy bin.}
 \label{fig:astroindependentexobounds}
\end{figure}

We compare these predictions to the 90\% C.L. bound coming from LUX for the efficiency corresponding to $\epsilon_{\mathrm{LY}}=0.8$.  Within each $\vmin$ bin we compare at the point where the LUX bound is weakest, for those where the lower bound on $\tilde{g}_{\mathrm{CDMS}}>\tilde{g}_{\mathrm{LUX}}$ we calculate the distance between the two results in terms of the $1\sigma$ uncertainty on the CDMS result.  The outcome of this comparison are shown in Figure~\ref{fig:astroindependentexobounds}.  From this we see that allowing for any physically allowed velocity distribution does not greatly relieve the tension.  However, if the highest energy event in CDMS was due to background the two 90\% C.L. results of CDMS and LUX can be made consistent with one another.  In addition if the light yield is less than anticipated the tension, as expected, is further reduced although for much of the parameter space even a reduction to $\epsilon_{\mathrm{LY}}=0.2$ will not be sufficient.  Furthermore, the distribution of events at CDMS, interpreted as being due to exothermic DM, seems to prefer a velocity distribution that is very different from the canonical MB distribution.

\subsection{Signals at CDMS}
As we have seen, it is very hard to find consistency between LUX and CDMS-Si such that all three events could arise from DM, unless isospin dependence suppresses the signal via a cancellation, even with very strong assumptions about the properties of LXe. However, with strong assumptions about \leff\ below 3 keV and an additional electric field dependent suppression \ely, elastic and exothermic scenarios are still possible. In these cases, we should also inquire about what the signals expected at CDMS will be.

\begin{figure}[t] 
   \centering
 \includegraphics[width=0.32\columnwidth]{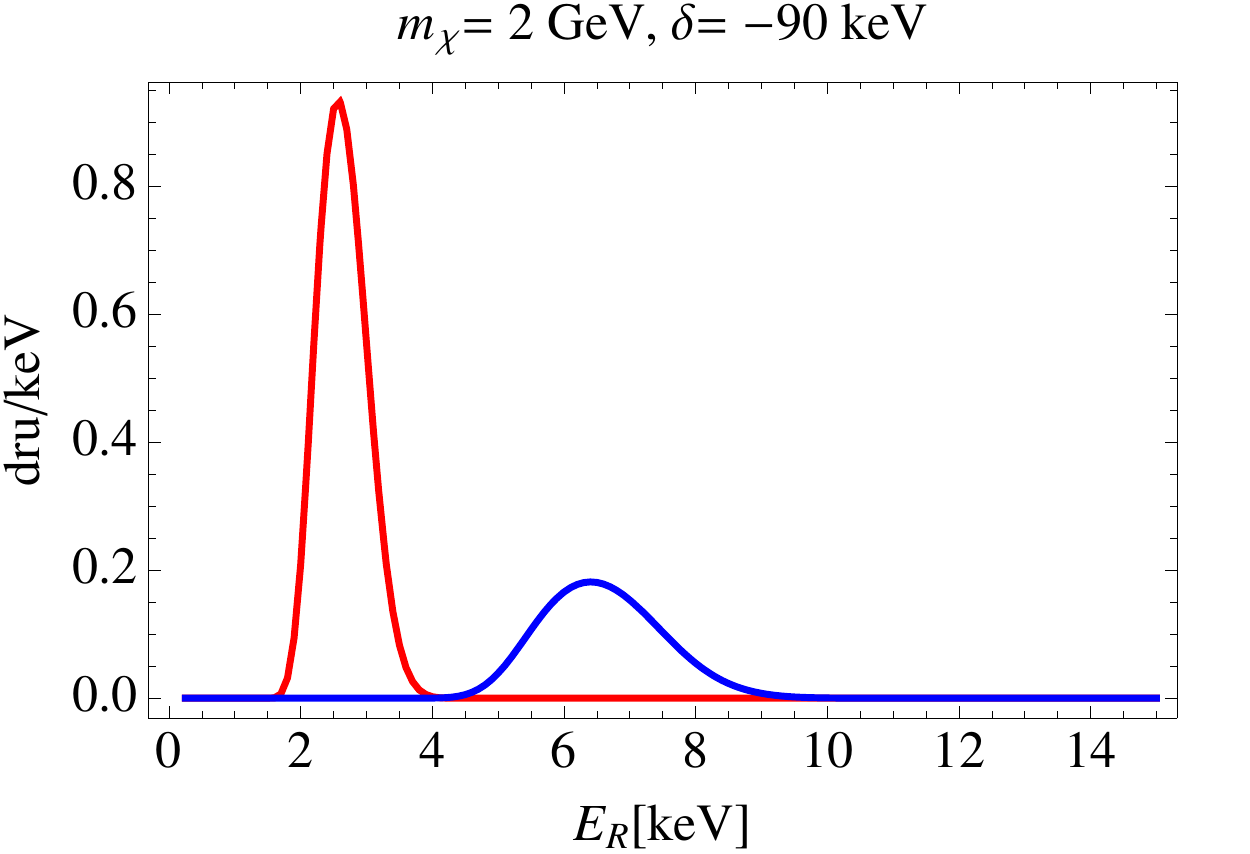} 
\includegraphics[width=0.32\columnwidth]{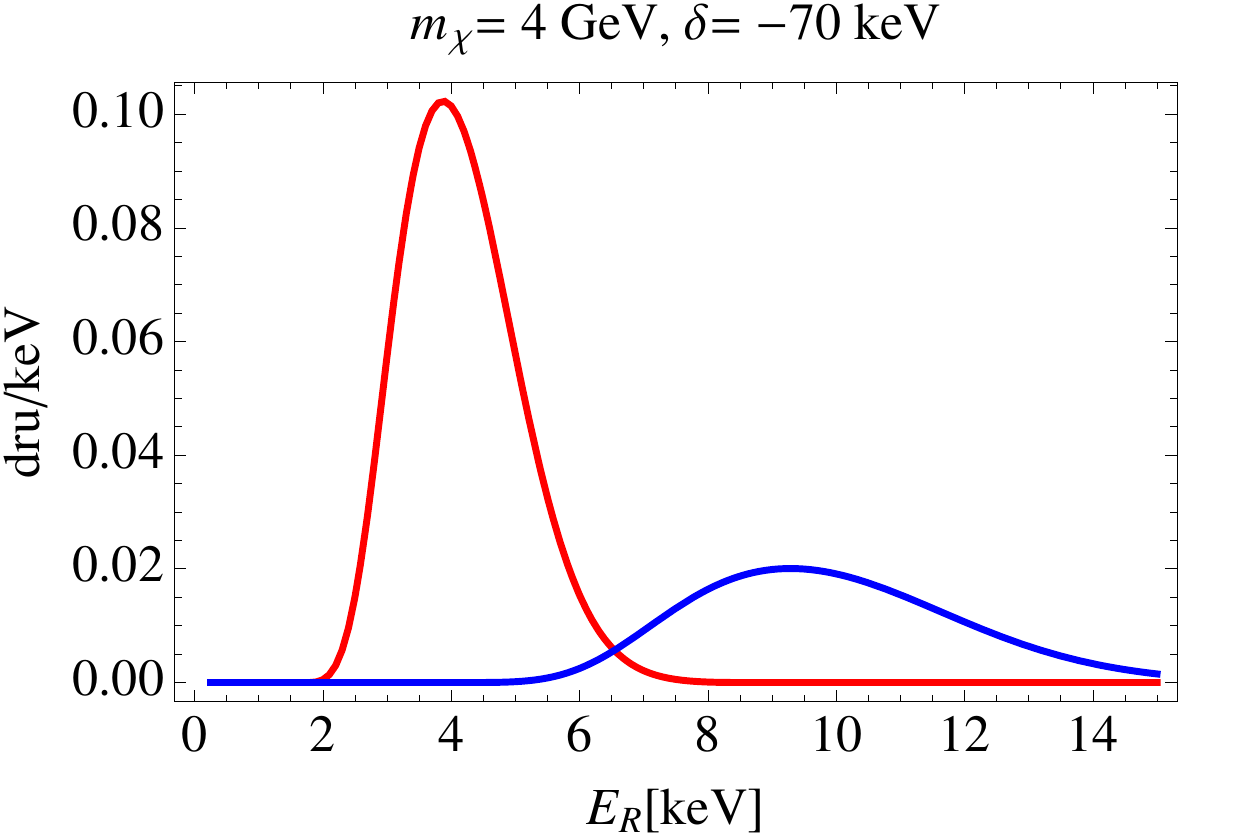} 
\includegraphics[width=0.32\columnwidth]{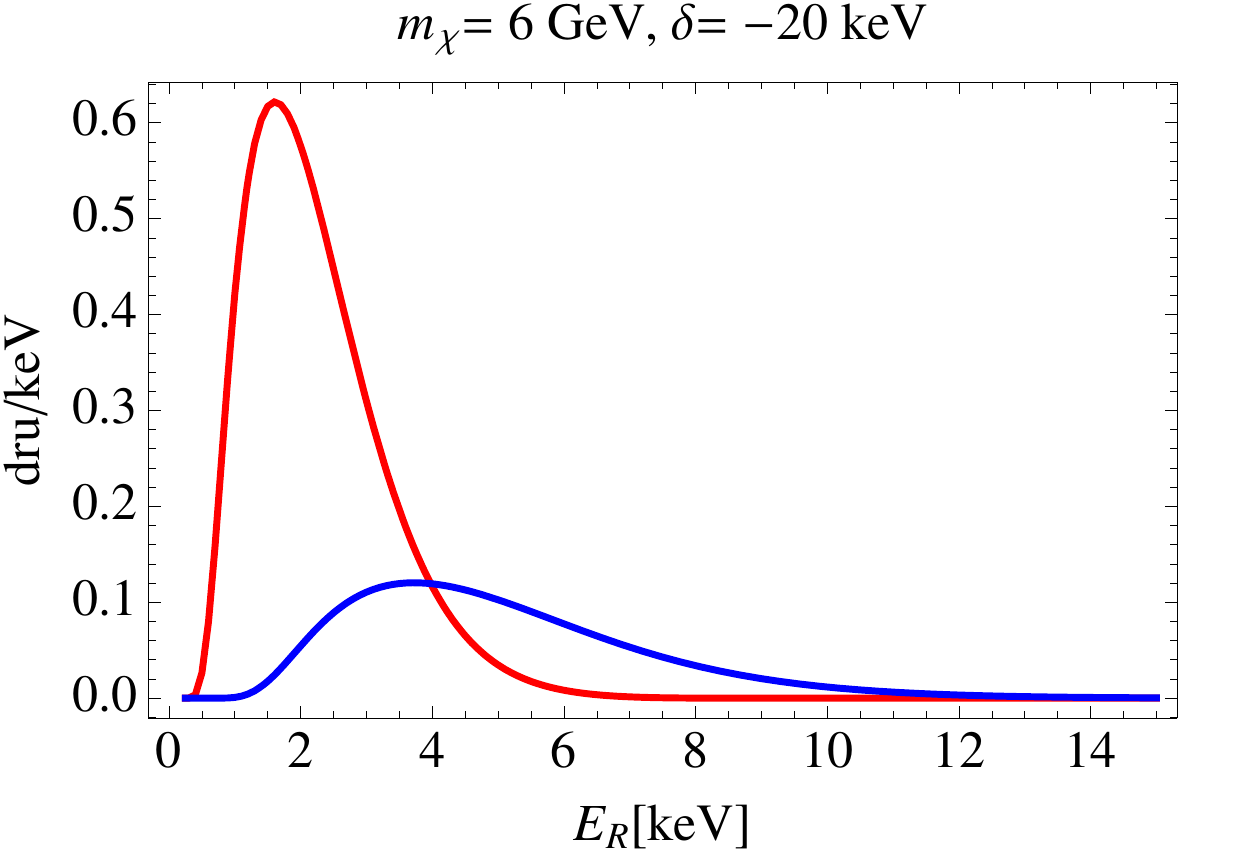} 
\caption{Signals at CDMS for three representative points. The expected rates for Si (Ge) in the range 5-6/6-7 keV (2+ keV/3+ keV) are 0.1/0.2 (0.9/0.2), 0/0 (0.2/0.2) and 0.09/0.06 (0.7/0.3) cpd/kg.}
 \label{fig:cdmssignals}
\end{figure}

We see that in not all cases is it clear what the signal should be below threshold for updated silicon analyses. For instance, while going just below the 7 keV threshold for a 2 GeV WIMP should show a sizeable signal at CDMS-Si, for 4 GeV it is essentially zero, and again for 6 GeV is sizable, but significantly smaller than the very light case. In contrast, all exothermic scenarios will produce a sizable signal above 2 and 3 keV for germanium, which should be testable at the upcoming SuperCDMS runs, if backgrounds remain low enough.

\section{Conclusions}

To investigate the light WIMP hypothesis, we have a large set of knobs to simultaneously consider. On the experimental front, we have at least three: $E_0$, the energy scale at which we assume \leff\  abruptly goes to zero (via a $\theta$ function), \ely, the overall suppression of light in the presence of the LUX electric field compared to background, and $N_{evt}$, the number of CDMS-Si events one actually wants to explain via dark matter. On the astrophysics side, there are the various parameters of the standard halo model, but also one can consider significant deviations from it. On the particle physics front, there is: $m_{\chi}$, which changes the relative sensitivity of different experiments, $f_p$ versus $f_n$, allowing one to adjust the sensitivities of various isotopes, and $\delta$, the inelasticity parameter, which can dramatically alter the kinematics of the process. 

Together this is a large set of knobs to turn simultaneously, and not all of them are simultaneously relevant. For instance, exothermic scattering tends to be largely insensitive to the tails of the velocity distribution, making deviations from the SHM less important. Some regions of parameter space are uninteresting - for elastically scattering WIMPs, for instance, while regions of compatible parameter space might open up at $m_\chi\sim 2\ \gev$, this is well below $\sim 4\ \gev$, where we would not expect any particles to have adequate kinetic energy to scatter (elastically) -- even for dramatic departures from the SHM.

Given the large number of permutations to consider, it is, not surprisingly, difficult to summarize them. However, there does seem to be a simple conclusion: if one wants to believe that all three events at CDMS-Si are due to dark matter, almost all scenarios require that the response of liquid xenon to nuclear recoils is dramatically different from what has previously been assumed or there is an extreme tuning of parameters. Even if only two events are from dark matter, the only scenario that does {\em not} require significant assumptions about the detector response is for an isospin-dependent interaction, where a $\sim 10\%$ tuning can reconcile the experiments.

Let us discuss this in more detail. We break the situations down first by $N_{evt}$. Assuming all three events are arising from dark matter, even assuming LXe has no scintillation below 3 keV (i.e., $E_0 = 3\ \kev$), for standard elastically scattering WIMPs, we would require $m_\chi \lsim 3 \gev$ for consistency. In such a mass range, we would require velocities well above the galactic escape velocity to yield the signals at CDMS. Even assuming a dramatic suppression in the light yield as a consequence of the electric field in LUX, this conclusion does not qualitatively change.  For instance, for 
$m_\chi\sim 5\ \gev$ we expect $\sim 100$ events at LUX with $\ely = 0.8$, and only $\sim 20$ with $\ely = 0.2$. If $E_0$ is below 3 keV, this only becomes more constrained. 

For exothermic scattering, if $\leff$\ drops suddenly at 3 keV, there is viable parameter space for $2\ \gev \lsim m_\chi  \lsim 4\ \gev$. However, this goes away if \leff\ is smoothly dropping below 3 keV and only abruptly goes to zero at 2 keV. In such a case, one must assume a dramatic suppression of \ely\ to 0.2 to find viable parameter space, in roughly the same mass range. 

For isospin dependent scattering, there is the standard highly tuned region around $f_n/f_p =-0.7$, extending up at roughly 7 GeV, even with $E_0 = 2\ \kev$ and $\ely = 0.8$. Thus, the three events can be understood with the standard assumptions about LXe, but at the expense of a significant tuning of the model parameter space.

If only two events at CDMS (specifically, the two lower energy events) are from DM, the situation changes, but not dramatically. For elastically scattering WIMPs, there is some allowed parameter space for $m_\chi \lsim 7\ \gev$, but we still must assume that \leff\ drops dramatically below 3 keV, even for highly suppressed $\ely \sim 0.2$. For somewhat lighter WIMPs, $m_\chi = 4\ \gev$ for such highly suppressed \ely, the additional suppression of \leff\ is not necessary (that is, the suppressed light yield need not drop abruptly to zero - the field dependent suppression assumed at \ely = 0.2 is adequate). However, neither of these effects is necessarily expected.

For exothermic scattering, the situation is nearly identical to the case where we assume all three events are due to DM. Namely, if $E_0 = 3\ \kev$, regions of parameter space appear (between $2\ \gev \lsim m_\chi  \lsim 4\ \gev$ as before), but with $E_0 = 2\ \kev$ we must assume the dramatic \ely = 0.2 suppression for viable parameter space to appear.

For isospin dependent scattering, the situation is somewhat improved. The tuning required is more like $\sim 10\%$ at $m_\chi \sim 7\ \gev$ if $E_0 = 2\ \kev$. Assuming either \ely = 0.2 or $E_0 = 3\ \kev$ allows essentially untuned scenarios.

In summary - we find that within the framework we have studied, there is a severe tension between LUX and the DM interpretation of the CDMS-Si events, especially if one believes all three CDMS events are from DM, with only highly tuned scenarios or significant deviations from the assumptions about LXe allowing consistency. For DM to explain only two events at CDMS, one needs only moderate tuning of model parameters, or similar changes to the assumptions about LXe.

Going forward then, it seems clear that the situation will not be concluded by further running of LUX or other LXe experiments such as Xenon1T - the experimental results from the state of the art LXe detectors is already limited only by the uncertainty of the properties of the detector itself. Further studies to show the properties of \leff\ below 3 keV and in the presence of electric fields on the scintillation of LXe are the most important steps that can be taken to improve the significance of those results.

In contrast, even very weak assumptions about the properties of LXe push us into a fairly narrow model corridor. Either fairly light DM (5-7\ \gev), or isospin-dependent scattering ($m_\chi \sim 7\ \gev$) are possible, in which case the signal below threshold at CDMS should appear in future studies of the Si data or new low threshold Ge runs. For instance, assuming $\sim$ 5 ton-days of exposure at SuperCDMS a 7 GeV WIMP with isospin-dependent scattering would lead to at least 120 (80) events in germanium if the threshold was 2 (3) keV, \emph{before} (the presently unknown) detector efficiencies are taken into account, a 5 GeV WIMP with $f_n=f_p$ predicts over 890 (530) events.

For exothermic scattering, the signal is most consistent for fairly light WIMPs. However, in these cases, clear signals should appear at CDMS. In many cases, just pushing the CDMS-Si threshold lower can help exclude a number of scenarios, especially where the signal is expected to rise dramatically below it, such as heavier (7+ GeV) elastically scattering WIMPs and very light ($\sim 2\ \gev$) exothermically scattering WIMPs. All these scenarios however would produce a significant signal at SuperCDMS, and should be visible with suitable background suppression. In particular, the exothermic scenarios often yield a peaked signal which may be easier to separate from background than standard exponentials.

While our conclusion is fairly pessimistic about the viability of these explanations, it is important to recognize that these conclusions are specifically for the models we have considered. Alternative scattering, such as via dipoles \cite{Gresham:2013mua} can modify the sensitivity, for instance, or more dramatic departures from the standard scattering hypothesis, such as \cite{Feldstein:2010su,Pospelov:2013nea} may appear that explain the differing experimental results quite straightforwardly. It is important to keep an open mind in these directions.

At the same time, future results by both LXe experiments as well as the CDMS collaboration will help clarify the situation, and whether such excursions from a standard WIMP are the only remaining possibility.

\section*{Acknowledgements} We thank Prateek Agrawal, Dan Hooper, and Felix Yu for helpful discussions.   
NW is supported by NSF grant PHY- 0947827 and PHY-1316753.  Fermilab is operated by Fermi
Research Alliance, LLC, under Contract DE-AC02-07CH11359 with the United States
Department of Energy. PF and NW would like to thank the KITP where part of this work was completed.  This research was supported in part by the National Science Foundation under Grant No. NSF PHY11-25915.  PJF and NW acknowledge partial support from the European Union FP7 ITN INVISIBLES (Marie Curie Actions, PITN- GA-2011- 289442). 

\bibliography{CDMSSi}

\end{document}